\documentclass[sigconf,nonacm,screen]{acmart}
\pdfoutput=1
\usepackage{tabularray}
\usepackage[noend]{algpseudocode}
\usepackage{multirow}
\usepackage{xurl}
\usepackage{booktabs} % For formal tables
\usepackage{amsmath}
\usepackage{graphicx}
\usepackage[belowskip=-12pt]{caption}
\usepackage{subcaption}
\usepackage{rotating}
\usepackage{tablefootnote}
\usepackage{adjustbox}
\usepackage{array}
\usepackage{float}
\usepackage{tabularx}
\usepackage{xcolor}
\usepackage{stfloats}
\usepackage{wrapfig}
\usepackage{listings}

\usepackage{color, colortbl}
\usepackage{balance} % to balance the bibliography
\usepackage{lscape}
\usepackage{xspace}
\usepackage{framed}
\usepackage{syntax}
\usepackage{parcolumns}
\usepackage{pifont}
\usepackage{soul}
\usepackage[tikz]{bclogo}
\usepackage{enumitem}
\usepackage{soul}
\usepackage{makecell}
\usepackage[many]{tcolorbox}
\usepackage{cleveref}
\usepackage{tikzsymbols}
\usepackage{textcomp}

\crefformat{section}{\S#2#1#3} % see manual of cleveref, section 8.2.1
\crefformat{subsection}{\S#2#1#3}
\crefformat{subsubsection}{\S#2#1#3}

\definecolor{maroon}{cmyk}{0,0.87,0.68,0.00}
\definecolor{mygreen}{cmyk}{0.41,0.1,0.5,0}

\newcolumntype{R}[2]{%
    >{\adjustbox{angle=#1,lap=\width-(#2)}\bgroup}%
    l%
    <{\egroup}%
}
\usepackage[linesnumbered,ruled,vlined]{algorithm2e}
\SetKwInput{KwInputs}{Inputs}                % Set the Input
\SetKwInput{KwOutputs}{Outputs}              % set the Output
\SetKwInput{KwOutput}{Output}              % set the Output

\definecolor{patchgreen}{HTML}{ccffcc}
\definecolor{patchred}{HTML}{ffcccc}
\definecolor{myyellow}{HTML}{FFFFCC}
\definecolor{mygray}{HTML}{F0F0F0}

\newcommand{\addgreen}{\makebox[0pt][l]{\color{patchgreen}\rule[-1ex]{\linewidth}{3ex}}}
\newcommand{\addred}{\makebox[0pt][l]{\color{patchred}\rule[-1ex]{\linewidth}{3ex}}}

\newcommand{\yang}[1]{\textcolor{blue}{[#1]}}

\newcommand\printpercent[2]{\the\numexpr#1*100/#2\%}

\preto\tabular{\setcounter{rownumbers}{0}}
\newcounter{rownumbers}

\newcommand{\approach}{\textsc{FlakyDoctor}\xspace}
\newcommand{\ifixflakies}{iFixFlakies\xspace}
\newcommand{\ipflakies}{iPFlakies\xspace}
\newcommand{\idflakies}{iDFlakies\xspace}
\newcommand{\odrepair}{ODRepair\xspace}
\newcommand{\idoft}{IDoFT\xspace}
\newcommand{\nondex}{NonDex\xspace}
\newcommand{\dexfix}{DexFix\xspace}
\newcommand{\inspector}{\textit{Inspector}\xspace}
\newcommand{\promptgeneration}{\textit{Prompt Generator}\xspace}
\newcommand{\repair}{\textit{Tailor}\xspace}
\newcommand{\validation}{\textit{Validator}\xspace}
\newcommand{\feedback}{\textit{Feedback Loop}\xspace}
\newcommand{\stitch}{\textit{Stitching}\xspace}
\newcommand{\gpt}{GPT-4\xspace}
\newcommand{\magic}{Magicoder\xspace}

\definecolor{problemblue}{RGB}{100,134,158}
\definecolor{idiomsgreen}{RGB}{0,162,0}
\definecolor{exercisebgblue}{rgb}{0,  .69,  .941}
\definecolor{deepgreen}{rgb}{0.0, 0.5, 0.0}
\definecolor{codegreen}{rgb}{0,0.6,0}
\definecolor{codegray}{rgb}{0.5,0.5,0.5}
\definecolor{codepurple}{rgb}{0.58,0,0.82}
\definecolor{backcolour}{rgb}{0.95,0.95,0.92}
\definecolor{redColor}{RGB}{255,0,0}
\definecolor{Gray}{gray}{0.1}
\definecolor{CadetBlue}{RGB}{243.0,42.1,134.0}

\newtcbtheorem[]{Theorem}{}{
        enhanced,
        sharp corners,
        colback=white,
        colframe=exercisebgblue
    }{thm}

\lstdefinestyle{code}{
  backgroundcolor=\color{gray!4},
  commentstyle=\color{codegray},
  keywordstyle=\color{codepurple},
  numberstyle=\tiny\color{codegray},
  stringstyle=\color{codegray},
  basicstyle=\ttfamily\footnotesize,
  breakatwhitespace=false,         
  breaklines=true,                 
  captionpos=b,                    
  keepspaces=true,                 
  numbers=left,                    
  numbersep=5pt,                  
  showspaces=false,                
  showstringspaces=false,
  showtabs=false,                  
  tabsize=2,
}
\lstdefinelanguage{test}{%
	language     = python,
	breaklines = true,backgroundcolor=\color{white},escapechar=!,rulecolor=\color{black}, breaklines=true,sensitive=true,  numbersep=5pt, xleftmargin=.015\textwidth, frame=tb,label=test
}
\lstdefinelanguage{source}{%
	language     = python,
	breaklines = true,
firstnumber=0,numberfirstline=false,columns=fullflexible,numbers=left,backgroundcolor=\color{white},
    rulecolor=\color{black}, 
    breaklines=true,sensitive=true, numbersep=5pt, xleftmargin=.015\textwidth, label=test
}

\definecolor{diffstart}{named}{codegreen}
\definecolor{diffincl}{named}{redColor}
\DeclareCaptionFormat{listing}{#3}
\newcounter{NumObservations}
\stepcounter{NumObservations}
\definecolor{shadecolor}{rgb}{.9,.9,.9}

\definecolor{msftBlue}{RGB}{0,164,239}
\definecolor{msftGreen}{RGB}{127,186,0}
\definecolor{msftYello}{RGB}{255,185,0}
\usepackage{bigstrut}
\definecolor{vlcolor}{rgb}{0.9,0.1,0.1}

\begin{document}

\title[A Generic Approach to Fix Test Flakiness in Real-World Projects]{A Generic Approach to Fix Test Flakiness in Real-World Projects}

\author{Yang Chen}
\affiliation{
    \institution{University of Illinois Urbana-Champaign, USA}
    \country{}
}
\email{yangc9@illinois.edu}

\author{Reyhaneh Jabbarvand}
\affiliation{
    \institution{University of Illinois Urbana-Champaign, USA}
    \country{}
}
\email{reyhaneh@illinois.edu}

\begin{abstract}
Test flakiness, a non-deterministic behavior of builds irrelevant to code changes, is a major and continuing impediment to delivering reliable software. The very few techniques for the automated repair of test flakiness are specifically crafted to repair either Order-Dependent (OD) or Implementation-Dependent (ID) flakiness. They are also all symbolic approaches, i.e., leverage program analysis to detect and repair \textit{known test flakiness patterns and root causes}, failing to generalize. To bridge the gap, we propose \approach, a neuro-symbolic technique that combines the power of LLMs---generalizability---and program analysis---soundness---to fix different types of test flakiness. 

Our extensive evaluation using $873$ confirmed flaky tests ($332$ OD and $541$ ID) from $243$ real-world projects demonstrates the ability of \approach in repairing flakiness, achieving $57$\% (OD) and $59$\% (ID) success rate. 
Comparing to three alternative flakiness repair approaches, \approach can repair $8\%$ more ID tests than \dexfix, $12\%$ more OD flaky tests than \odrepair, and $17\%$ more OD flaky tests than \ifixflakies. Regardless of underlying LLM, the non-LLM components of \approach contribute to $12$--$31$ $\%$ of the overall performance, i.e., while part of the \approach power is from using LLMs, they are not good enough to repair flaky tests in real-world projects alone. 
What makes the proposed technique superior to related research on test flakiness mitigation specifically and program repair, in general, is repairing $79$ \textit{previously unfixed} flaky tests in \textit{real-world projects}. We opened pull requests for all cases with corresponding patches; $ 19$ of them were accepted and merged at the time of submission. 

\end{abstract}
%-----------------------------------------------



% \ccsdesc[500]{Software and its engineering~Software maintenance tools}
% \ccsdesc[300]{Software and its engineering~Software evolution}
% \ccsdesc[300]{Computing methodologies~Machine learning}
% \ccsdesc[300]{Software and its engineering~Domain specific languages}


% \keywords{}

\maketitle

\section{Introduction}
\label{sec:introduction}

Test flakiness is the problem of observing non-determinism in test execution results without any changes in the code under tests. This phenomenon can drastically impact the effectiveness of regression testing and software products. The root cause of test flakiness is code smells in the test suite. However, developers cannot distinguish if the test failure is due to a bug in the code or flakiness, which can waste the valuable time of developers~\cite{eck2019understanding} and computing resources~\cite{kowalczyk2020modeling,micco2017state} without resolving the underlying issue. 

To minimize the negative impact of test flakiness, several techniques have been proposed to characterize,
%~\cite{lam2019root,lam2020study,lam2020large,lam2020understanding,dutta2020detecting,chen2023transforming}, 
detect,
%~\cite{ziftci2020flake,wang2022ipflakies,person2015test,mascheroni2018identifying,king2018towards,pinto2020vocabulary,bell2018deflaker,lam2019idflakies,wei2022preempting,verdecchia2021know,yi2021finding}, 
and repair
%~\cite{dutta2021flex,li2022repairing,pei2023traf,shi2019ifixflakies,wang2022ipflakies} 
them. Compared to detecting flakiness, there is a dearth of work focusing on their repair. All such techniques repair a specific type of test flakiness. For example, \ifixflakies~\cite{shi2019ifixflakies}, \ipflakies~\cite{wang2022ipflakies}, and \odrepair~\cite{li2022repairing} are all designed to repair Order-Dependent (OD) flakiness, which non-deterministically pass or fail under different test execution orders. \dexfix~\cite{zhang2021domain} proposes a set of domain-specific strategies to repair Implementation-Dependent (ID) flaky tests, which happen due to unrealistic assumptions about non-ordered collections. TRaF~\cite{pei2023traf} repairs asynchronous waits, a specific type of Non-Order-Dependent (NOD) tests that non-deterministically pass or fail due to concurrency issues or dependency on the execution platform, memory, and time.

%Non-Order-Dependent (NOD) flaky tests are another category that non-deterministically pass or fail due to concurrency issues or dependency on the execution platform, memory, and time. Except for TRaF~\cite{pei2023traf}, which repairs a specific type of NOD tests, i.e., asynchronous waits, no other work aimed to repair NOD flaky tests. There is also no automated technique for repairing Non-Idempotent-Outcome (NIO) flaky tests~\cite{wei2022preempting}, which pollute their states and manifest under multiple executions of the same test order. 

%one paragraph explaining the proposed technique
Regardless of the flakiness category of interest, all prior techniques are symbolic,
%\footnote{Please note that the keyword symbolic here refers to a general term of symbolic learning in contrast to machine learning and should not be confused with symbolic execution.}, 
i.e., they use human knowledge to devise and implement analytical rules for repairing test flakiness. As a result, they cannot generalize to repairing flaky tests with unknown root causes that analytical rules do not implement. More importantly, their abilities are limited due to the potential limitation of underlying program analysis techniques in generalizing to new programming features and various development styles. 

Large Language Models (LLMs) are effective in generative programming tasks, making them a natural solution for overcoming the generalizability limitations of fixing flaky tests.
%~\cite{feng2020codebert,guo2020graphcodebert,wang2021codet5,wang2023codet5+,nijkamp2022codegen,fried2022incoder,ahmad2021unified,chakraborty2022natgen,jin2023inferfix,pan2023understanding,wang2023leti,xia2023conversational}. A natural solution to overcome the generalizability limitations for fixing flaky tests is using learning-based automation, especially, using LLMs.
However, LLMs also suffer from a series of limitations, namely, (L1) generating (syntactically and semantically) incorrect code~\cite{liu2023your,ouyang2023llm,liu2024codemind}, (L2) the need for proper context in the prompt to perform reasonably~\cite{white2023prompt,pan2023understanding,shrivastava2023repository}, and (L3) limited context window, which makes leveraging them for real-world programs and test suites challenging~\cite{liu2023your,ouyang2023llm,pan2023understanding}. To use LLMs for fixing flakiness in real-world problems, one can mitigate these challenges by augmenting LLMs with sound symbolic approaches to resolve syntactic issues and validate the generated code (L1), and extract the \textit{minimum} amount of \textit{relevant context} for the prompt to achieve the best possible result (L2 and L3).   

We propose \approach, a neuro-symbolic approach that combines the generalizability power of LLMs with the soundness of program analysis for repairing OD and ID flaky tests. \approach takes the name and type of flaky test as input (\cref{subsec:inspector}) and extracts its test code and body of other tests that partnered in crime. It then executes them and localizes the source of flakiness. By including the above information as problem context to generate a prompt (\cref{subsec:promptgeneration}), it instructs LLMs to create a patch for repairing flakiness (\cref{subsec:repair}). If the patch has compilation errors, \approach first tries to solve the compilation issues offline and then forwards the patch for validation if resolved (\cref{subsec:validation}). It terminates with success if the validation confirms the generated patch resolves the flakiness. Otherwise, 
%inspired by prior work that shows iterative, feedback-based prompting of LLMs improve the overall performance~\cite{xu2023reprompting,wang2023leti,chen2022codet}, 
\approach updates the prompt with \textit{concise} information about unresolved issues and makes subsequent repair attempts (\cref{subsec:feedback}). Repairing terminates after a fixed number of iterations or when all the flaky tests are repaired. 
Our notable contributions are:

\begin{itemize}[leftmargin=*]
    \item \textbf{Technique.} To our knowledge, \approach is the first technique for repairing more than one category of test flakiness. Prior work focused on repairing one type of test flakiness, OD flaky tests~\cite{shi2019ifixflakies,wang2022ipflakies,li2022repairing} or ID flaky tests~\cite{zhang2021domain}. Also, none of the prior techniques has leveraged the power of LLMs in repairing test flakiness. The power of \approach is not directly from the underlying LLM: offline fixing of issues and precise bug localization by minimizing the amount of feedback using static analysis contributes to $12$--$31$ $\%$ of its performance, depending on the underlying LLM. \approach 
    %is publicly available~\cite{website} and
    works with both API- and open-access LLMs.

    \item \textbf{Evaluation} We comprehensively evaluated the effectiveness of \approach in repairing $873$ flaky tests from $243$ real-world projects. Our empirical results corroborate the ability of \approach in repairing $58$\% of studied flakiness ($57\%$ OD and $59\%$ ID) in $103$ seconds, on average. Among the correct patches, $79$ of them were not previously fixed by developers or any existing automated techniques. We opened PRs for those repaired flaky tests, and $19$ of them were accepted and merged by the time of submission.
    %\footnote{Given that reviewing PRs takes time, we will update the number of accepted PRs on the artifact website for reviewers to check}. 
    Compared to alternative approaches, \approach can repair $8\%$ more ID tests than \dexfix, $12\%$ more OD flaky tests than \odrepair, and $17\%$ more OD flaky tests than \ifixflakies.

    \vspace{-10pt}

    % \yang{do we report the aggregated time cost here? because I calculate the time cost for successful/unsuccessful patches separately in RQ.}

    %\item \textbf{Tool.} We implemented \approach as a publicly available tool~\cite{website} for the research community and practitioners to repair flaky tests. It works with both API- and open-access LLMs. 
    
\end{itemize}

\section{Background and Motivation}
\label{sec:background}
Depending on whether changing the execution order plays a part in manifesting test flakiness, prior research categorizes flaky tests into OD and NOD. This section explains these categories with real-world examples. We also discuss potential challenges in repairing different types of flaky tests and explain why \approach could repair flakiness in the examples, while alternative approaches failed to patch them. 

\textbf{OD Flaky Tests.} 
Such flakiness occurs when two or more tests in the test suite are coupled through a shared state that the developers do not properly manage, e.g., in \texttt{\small{tearDown}} or \texttt{\small{setUp}} methods~\cite{zhang2014empirically}. Test prioritization~\cite{rothermel1999test} or test parallelization~\cite{candido2017test} can change the execution order of the tests, altering their outcome from pass to fail or vice versa. Tests that change the outcome due to polluted shared status are called \textit{victim} or \textit{brittle}~\cite{shi2019ifixflakies}. Victim tests pass when executed alone (but can fail if executed after some other tests), while brittle tests fail when run alone (but can pass when run after some other tests). A test that changes the shared state for the \textit{victim} test is called \textit{polluter}, while the test that changes the shared state for the \textit{brittle} is called \textit{state-setter}. 
%A victim test passes if executed before the polluter and fails otherwise. In contrast, a brittle test fails if executed before the state-setter and passes otherwise. 
In addition to polluter/victim and state-setter/brittle tests, \textit{cleaners}~\cite{shi2019ifixflakies} and \textit{state-unsetters}~\cite{chen2023transforming} are also important concepts related to OD flaky tests. When a cleaner test runs between a polluter and a victim, it cleans the polluted state so the victim can pass. Likewise, when a state-unsetter runs between a state-setter and a brittle, it neutralizes the state change impact, and the brittle fails. 

\begin{figure}[t]
\centering
\vspace{-5pt}
\noindent\begin{minipage}{.97\linewidth}{
\begin{lstlisting}[language = java,basicstyle=\small, style=code,numbersep=5pt,showlines=true,columns=fullflexible,escapechar=|]
// OD-Polluter
@Test 
public void assertGetEventTraceRdbConfigurationMap() {
  Properties properties = new Properties();  
  properties.setProperty(BootstrapEnvironment.EVENT_TRACE_RDB_DRIVER, "org.h2.Driver");
  properties.setProperty(BootstrapEnvironment.EVENT_TRACE_RDB_URL, "jdbc:h2:mem:job_event_trace");
  properties.setProperty(BootstrapEnvironment.EVENT_TRACE_RDB_USERNAME, "sa");
  properties.setProperty(BootstrapEnvironment.EVENT_TRACE_RDB_PASSWORD, "password");
  ReflectionUtils.setFieldValue(bootstrapEnvironment,"properties", properties); 
    //...
|\addgreen|+  ReflectionUtils.setFieldValue(bootstrapEnvironment,
|\addgreen|+     "properties", new Properties());
}  
// OD-Cleaner (Does not exist in the original test suite and has been added for illustration)
@Test 
public void cleaner() {
    ReflectionUtils.setFieldValue( bootstrapEnvironment, "properties", new Properties());
}
// OD-Victim
@Test 
public void assertWithoutEventTraceRdbConfiguration(){
    assertFalse(bootstrapEnvironment.getTracingConfiguration().isPresent());
}
\end{lstlisting}}
\end{minipage}
\vspace{-5pt}
\caption{ Example of a previously unfixed OD flakiness in Elasticjob~\cite{shardingsphere-elasticjob} repaired by \approach that cannot be repaired by alternative approaches}
\label{fig:od-example}
\vspace{-6pt}
\end{figure}
%The code is slightly modified (without changing the logic) for readability purposes. For example, \texttt{\small{BootstrapEnvironment.EVENT_TRACE_RDB_DRIVER}} substitutes \texttt{\small{BootstrapEnvironment.{\color{red}EnvironmentArgument}.EVENT_TRACE_RDB_DRIVER.{\color{red}getKey()}}} API call in the original implementation
% \vspace{-15pt}

\sloppy Figure~\ref{fig:od-example} shows polluter and victim tests from Elasticjob project~\cite{shardingsphere-elasticjob}. 
The shared state causing the dependency is the \texttt{\small{bootstrapEnvironment}}, a global variable in the test class. If the polluter runs before the victim, it will alter the state of \texttt{\small{bootstrapEnvironment}} (Lines 4--9 in polluter code), which causes the assertion in Line 22 of the victim to fail. Otherwise, the victim test passes. The cleaner---which does not exist in reality and has been added for illustration---neutralizes the impact of polluters by resetting the shared state, resulting in the victim pass.  

\noindent \textbf{\textit{Repair Challenge.}} 
The obvious solution to repair the OD flakiness is to remove the dependency. In the illustrative example of Figure~\ref{fig:od-example}, the patch should reset the properties of \texttt{\small{bootstrapEnvironment}} at the beginning of victim or the end of the polluter. Without the privilege of code synthesis ability of LLMs, prior techniques rely on the existence of cleaners to extract specific statements to clean the pollution and generate patches~\cite{shi2019ifixflakies}. To alleviate this need, subsequent techniques automatically generate the cleaner first ~\cite{li2022repairing} and use it for patch generation. Those techniques are still limited to the ability of testing techniques to generate correct cleaners.% tests. 

\approach leverages checks the potential shared state/variables between tests (\cref{subsec:inspector}) and then instructs LLM to modify the code of polluter to clear the polluted state (\cref{subsec:promptgeneration}). The highlighted line at the end of the polluter (Line 11) shows the patch for this real-world example generated by \approach. \ifixflakies could not fix this flakiness due to the absence of cleaners in the test suite. \odrepair detected the shared state successfully but could not generate the cleaner test to use it for repair further. 
%NIO flaky tests share a similar challenge, namely how to clean the pollution caused by the test itself, while it doesn't rely on the status shared with others.

\textbf{NOD Flaky Test.} 
NOD flakiness happens due to misuse or misunderstanding of programming APIs, concurrency problems, execution platforms, runtime environment, etc. Compared to OD flakiness, NOD flakiness occurs for each test in isolation and regardless of test execution order. As a result, one can detect or validate the patch by re-executing it without shuffling the test order. Still, detecting or validating the patch for NOD flakiness is challenging when the probability of observing flaky behavior is tiny~\cite{chen2023transforming}.
%, e.g., test-flakiness detection tools may re-execute the test suite more than $1000$ times to make a passing test to fail~\cite{lam2020large}.  
A special sub-category of NOD tests is ID flaky tests, which occurs due to incorrect assumptions about the non-ordered collections ~\cite{zhang2021domain}. 

Figure~\ref{fig:id-example} illustrates an ID flaky test from Hadoop~\cite{hadoop}, which 
happens due to converting an unordered collection (a \texttt{\small{Json}} object) into \texttt{\small{String}}. This is not problematic unless we assume a specific order for an unordered collection.
%behaves similarly to an ordered collection. 
That is, in Lines 8--9, the assertion checks if the conversion equals to ``\{"A":6,"B":2,"C":2\}'', assuming that the string conversions of the same \texttt{\small{Json}} object are always similar. As a result, the execution of this test non-deterministically passes or fails.  
    
%\noindent \textbf{\textit{Repair Challenge.}} The first step in repairing ID tests is understanding the source of non-determinism, i.e., localizing the bug. One approach is leveraging static analysis and looking for code smells in the test suite. However, such a technique that is explored before~\cite{zhang2021domain} may not generalize well due to various root causes; only a subset of them has been explored~\cite{lam2020large,chen2023transforming}. In the illustrative example of Figure~\ref{fig:nod-example}, the flakiness is peculiar, and so is the repair created by the developers. The patch replaces Lines 17--20 with Lines 21--26, since \texttt{\small{PowerMockito}} is a more powerful mocking framework compared to \texttt{\small{Mockito}}. However, \texttt{PowerMockito} is deprecated and does not work with JDK 9 and above. Consequently, when the project upgrades to a new JDK version, the test can become flaky again (or fail deterministically). In general, repairing NOD tests could be complicated.

\noindent \textbf{\textit{Repair Challenge.}} The first step in repairing ID tests is understanding the source of non-determinism, i.e., localizing the bug.
%, which happens at the call site of specific programming APIs. 
The key idea in \approach is that test execution failure can help localize the source of flakiness systematically. To extract such information, \approach analyzes the stack trace of the test execution failure and identifies the tests and statements within them that incorporate non-determinism. It then instructs the LLM to focus on specific lines in the culprit tests to repair the issue. 

To repair the ID flakiness in Figure~\ref{fig:id-example}, the patch generated by \approach first transforms the converted \texttt{\small{JSon}} object (\texttt{\small{csv1}}) into a \texttt{\small{LinkedHashMap}} (Line 10--12), and then reconstructs the expected output in the previous assertion as a \texttt{LinkedHashMap} (Lines 13--16). Comparing these two objects in the new assertion (Lines 17--18) resolves the flakiness. We suspect the LLM component of \approach was able to reason about the return type of \texttt{\small{mbs.getAttribute}} being \texttt{\small{JSon}}, based on the format of \{"A":6,"B":2,"C":2\} in the assertion argument (Line 9). The change of data leakage is narrow since this ID test was previously unfixed. The alternative approach for fixing ID tests, \dexfix, could not patch this flakiness as it looks for particular patterns and explicit usages of unordered collections, missing this example.

\noindent In this paper, we only focus on repairing OD and ID flakiness. The main reason is that \textit{reliable} detection of NOD tests in general, hence validating the generated patches, is still an open problem. Previous research studies rely on re-executing test suites $100$--$1000$ times to detect NOD flakiness. Even with this large number of executions, not observing flaky behavior does not mean it does not exist. %Assuming that would greatly threaten the validity of the results, which motivated us to evaluate \approach on OD and ID tests.

\begin{comment}
\input{CodeExamples/nod-example}
 
Figure~\ref{fig:nod-example} shows a real-world NOD flakiness from project Secor~\cite{secor}. The source of flakiness in test \texttt{\small{testDelimitedTextFileWriter}} is from the flaky behavior of \texttt{\small{Mockito}}, used in the class helper method \texttt{\small{mockDelimitedTextFileWriter}} for mocking the FileSystem class. The underlying issue in the mocking framework results in occasional \texttt{\small{IllegalStateException}} thrown by Lines 17--20 of \texttt{\small{mockDelimitedTextFileWriter}}, called at Line 6 of the NOD flaky test. 
\end{comment}

\vspace{-5pt}

% \vspace{-5pt}
\section{\approach}
\label{sec:approach}

Figure~\ref{fig:overview} shows the overview of \approach, consisting of \textit{four} main components, namely, \inspector, \promptgeneration, \repair, and \validation. The \inspector takes a flaky test suite as input, analyzes the test execution results, and localizes the source of test failures in the test code. 
Depending on the type of flakiness, a combination of inspection results, culprit test method(s), and relevant global variables and helper methods in the test class will be used by \promptgeneration to create the prompt. The prompt specifically instructs LLM to focus on particular statements and generate the patch by modifying the provided tests, variables, and helpers. 

Once the LLM responds to the prompt with a patch, \repair first checks for compilation errors, which are inevitable in the code produced by LLMs. In case of compilation issues in the patch, the \repair tries to resolve them offline in \textit{Stitching} sub-component. If the modified code passes compilation, it goes to the \validation to check if it resolves the flakiness. If validated, \approach terminates successfully. Otherwise, the incorrect patch generated from the current iteration, along with concise compilation or test execution outputs, goes for another round of inspection and repair. 
The iterative repair terminates upon generating a successful patch or for a fixed number of iterations. We will explain the details of each component in the remainder of this section.

\begin{figure}[t]
\centering
\vspace{-5pt}
\noindent\begin{minipage}{.97\linewidth}{
\begin{lstlisting}[language = java,numbersep=5pt,showlines=true,columns=fullflexible,escapechar=|]
@Test // ID flaky test
public void testPriority() throws Exception {
   //...
    MBeanServer mbs = ManagementFactory.getPlatformMBeanServer();
    ObjectName mxbeanName = new ObjectName(
     "Hadoop:service="+ namespace + ",name=DecayRpcScheduler");
    String cvs1 = (String)mbs.getAttribute(mxbeanName,"CallVolumeSummary");
|\addred|-   assertTrue("Get expected JMX of CallVolumeSummary before
|\addred|-       decay", cvs1.equals("\{"A":6,\"B\":2,\"C\":2\}"));
|\addgreen|+   Map<String, Integer> map1 = new Gson().fromJson(
|\addgreen|+       cvs1, new TypeToken<LinkedHashMap<String, Integer>()
|\addgreen|+       .getType());
|\addgreen|+   Map<String, Integer> expectedMap1 = new LinkedHashMap<>();
|\addgreen|+   expectedMap1.put("A", 6);
|\addgreen|+   expectedMap1.put("B", 2);
|\addgreen|+   expectedMap1.put("C", 2);
|\addgreen|+   assertEquals("Get expected JMX of CallVolumeSummary before  
|\addgreen|+        decay", expectedMap1, map1);
}
\end{lstlisting}}
\end{minipage}
% \vspace{-5pt}
\caption{ Example of a previously unfixed ID flakiness in Hadoop~\cite{hadoop} repaired by \approach that cannot be repaired by alternative approaches}
\label{fig:id-example}
\vspace{-5pt}
\end{figure}

\begin{figure*}
    \centering
    % \vspace{-10pt}
    \includegraphics[width=0.85\textwidth]{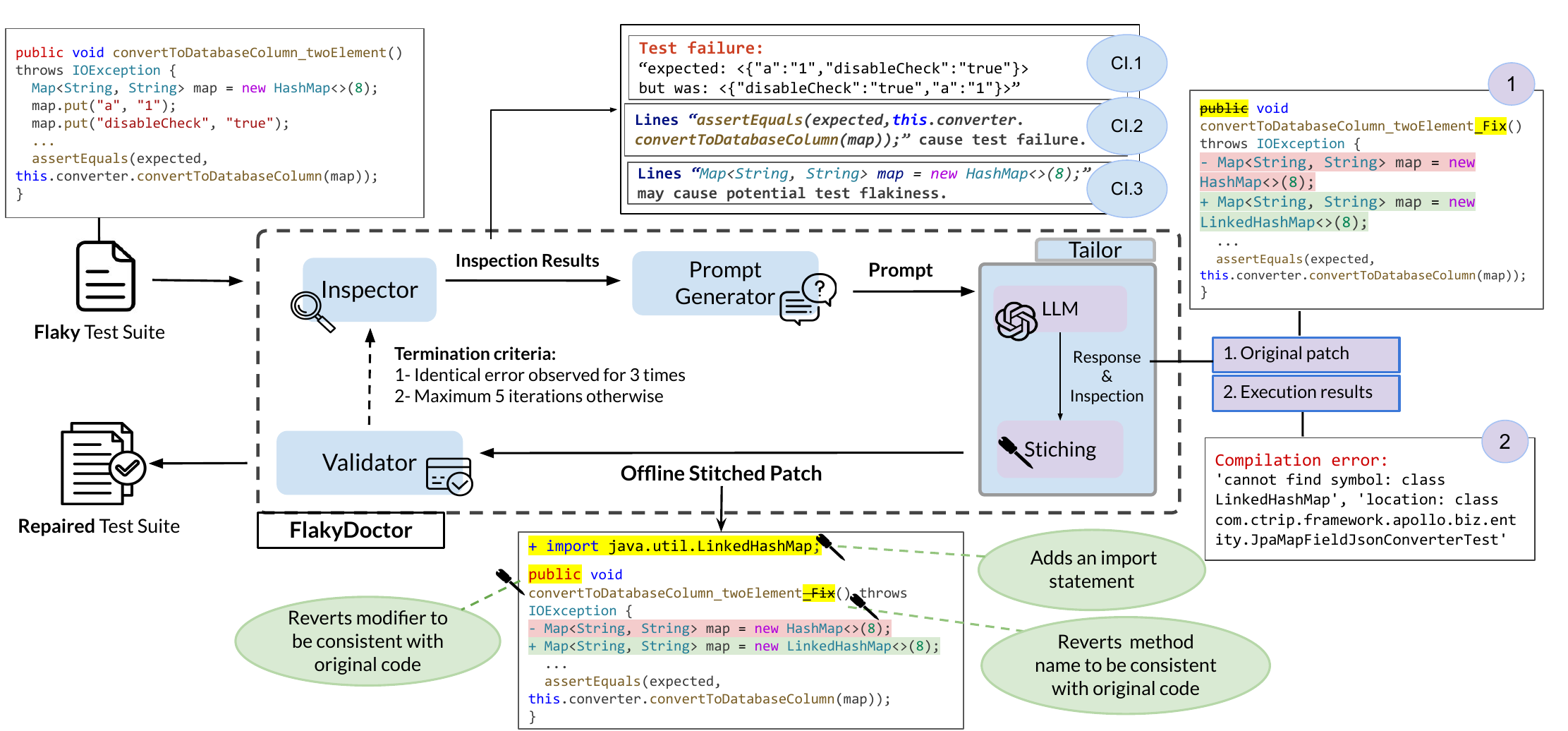}
    \caption{ Overview of \approach for repairing test flakiness 
    }
    \label{fig:overview}
\end{figure*}

\subsection{\inspector}
\label{subsec:inspector}

The \inspector takes the flaky test suite as input and extracts proper and concise contextual information required to repair the flakiness. If a given test suite contains more than one flaky test, the \approach will analyze them individually, and each will be patched separately. \inspector should generate three contextual information (CI) for the \promptgeneration: (CI.1) test execution errors, (CI.2) corresponding failed assertions, and (CI.3) potential source of flakiness. To that end, it first executes the tests to reproduce the failure. For OD-Victim tests, \inspector uses a modified version of Surefire~\cite{surefire} to specify the execution order of the polluter and victim, i.e., executes the polluter test before the victim to make it fail. For OD-Brittle tests, \approach executes them in isolation, as they fail by default. For ID tests, \inspector executes them with \nondex, which randomly explores different behaviors of certain APIs during test execution through multiple rounds to produce the failure outcome.  

After test execution and reproducing the failure, \inspector extracts the errors (CI.1) directly from the execution result. In the running example of Figure~\ref{fig:overview} that shows an ID flaky test \texttt{\small{convertToDatabaseColumn_twoElement}}, the error message (CI.1) is \texttt{\small{expected:\{"a":"1","disableCheck":"true"\} but was:\{"disableCheck":"true","a":"1"\}}}). By parsing the stack trace, \inspector can extract the line number in the test class and get the assert statement causing the failure accordingly (CI.2). 

Repairing without bug localization information is searching for a needle in a haystack. \inspector employs a method-level localization, i.e., only includes the flaky test methods instead of the entire test suite. It additionally employs the following heuristics to localize the source of flakiness at the statement level as much as possible:

\begin{itemize}[leftmargin=*]
    \item For ID flaky tests, it performs a flow-sensitive analysis to pinpoint unordered collections (e.g., \texttt{\small{HashMap}}) or APIs (e.g., \texttt{\small{getFields}}) related to the failed assertions, which may lead to a non-deterministic order of elements. In the running example, \inspector identifies \texttt{\small{Map<String, String> map = new HashMap<>(8)}} initializes an unordered collection and returns it as the potential cause of flakiness (CI.3).

    \item For OD-Victim tests, \inspector extracts global variables and helper methods such as \texttt{\small{setUp}} and \texttt{\small{tearDown}}. Global variables can be potential sources of dependency between tests. Including the helper methods is two-fold: they can be either a source of dependency between tests due to improper management of global variables and resources, or the patch can implement the fix inside them. 

\end{itemize}

\subsection{\promptgeneration}
\label{subsec:promptgeneration}

The current implementation of the \approach supports fixing ID and OD (OD-Victim and OD-Brittle) flaky tests and has three prompt templates corresponding to each type. Figures~\ref{fig:id-od-prompt} shows the templates for OD-Victim (Figures~\ref{fig:id-od-prompt}a) and ID (Figures~\ref{fig:id-od-prompt}b) flaky tests. The structure of prompt templates is similar, but \promptgeneration fills them differently according to flakiness type. 

The prompt starts with a natural language instruction, asking the LLM to repair test flakiness (Instruction section). If the LLM is instruction-tuned, the prompt asks it to act as a software testing expert to increase the chance of LLM producing a better response~\cite{gpt4}. Depending on the type of flakiness, the instructions provide more information specific to the flakiness type and general advice in repairing them. 

Next, the prompt introduces the problem that LLM should solve, i.e., repairing flakiness, by listing the names of the tests involved (Problem Definition), followed by relevant source code (Related Code) extracted by \inspector (CI.2). For ID and OD-Brittle flaky tests, the Related Code section only includes the flaky test declaration and implementation. For OD-Victim tests, this section includes the code of the victim, polluter, global variables, and helper methods. With this design decision, repairing a polluter or helper methods may resolve several other related flakiness in the test suite. \promptgeneration also concatenates statements that raise errors/failures (CI.1) and potential sources of flakiness (CI.2) to the prompt (Failure Location section) to help models localize the flakiness better, and, hopefully, generate a higher-quality patch.

\begin{figure}
    \centering
    % \vspace{-5pt}
    \includegraphics[width=0.5\textwidth]{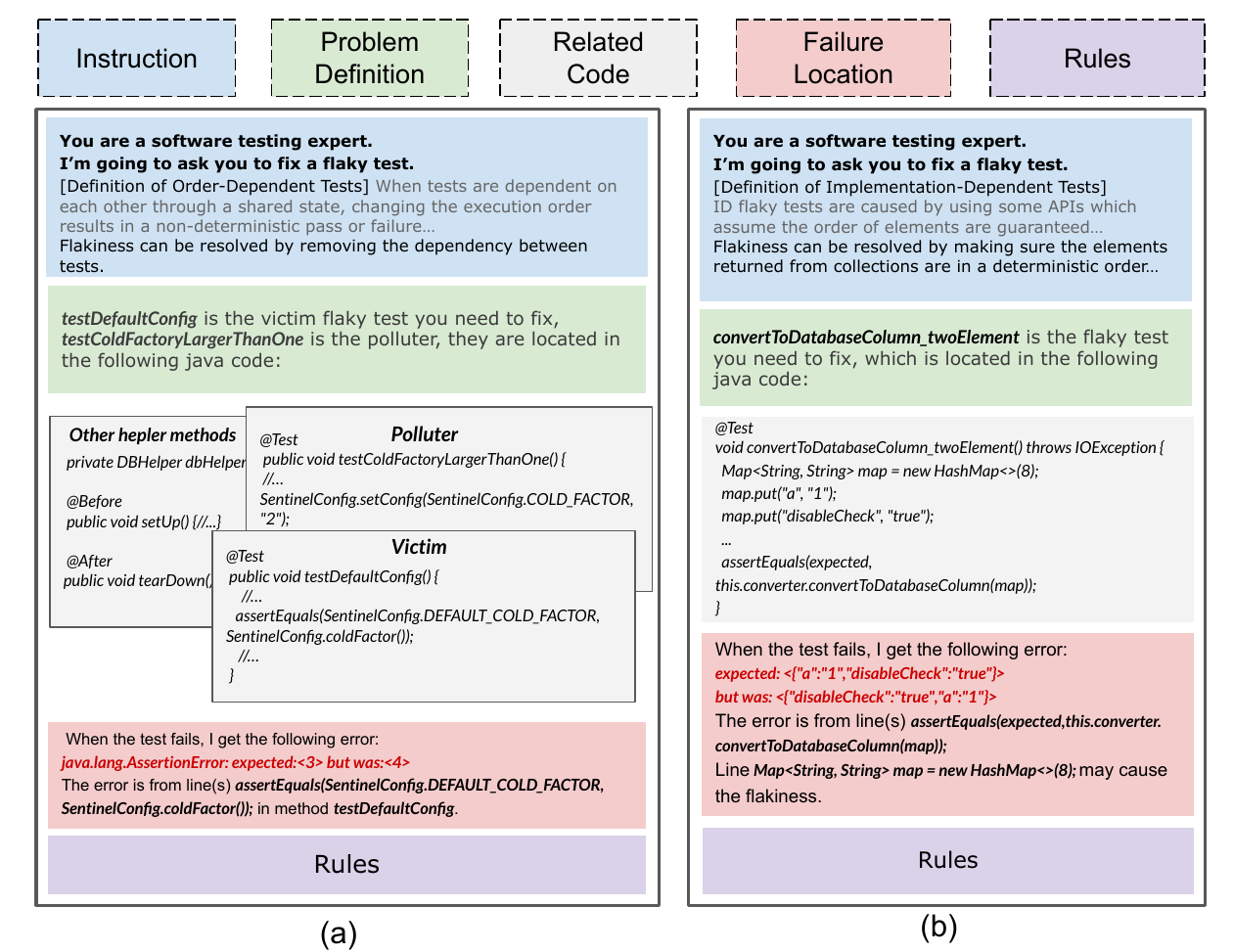}
    \caption{Prompt templates for repairing OD-Victim (a) and ID (b) test flakiness 
    }
    \label{fig:id-od-prompt}
    % \vspace{-5pt}
\end{figure}

\promptgeneration concludes the prompt with a list of \textit{six} rules for LLM to follow: (1) Solve the problem with implicit Chain-of-Thoughts (CoT)~\cite{wei2022chain}, (2) Update the imports and build files if needed, (3) Generate syntactically correct code, (4) Ensure all the arguments are correct, (5) Use compatible types for all variables, and (6) Follow the specified formatting (to facilitate response processing). As we will show later, including these rules helps only to a limited extent, which requires additional effort to compensate for the subpar performance of LLMs~\cite{ouyang2023llm}.
%\reyhan{please update the list with all the rules. In Figure 4, remove the enumerated list and demonstrate it with a placeholder such as \$rules}

\subsection{\repair}
\label{subsec:repair}

\repair consists of two sub-components: LLM and \stitch. The LLM carries most of the repair burden. \approach can work with any LLM with minimal changes in the prompts, and its current implementation uses GPT-4~\cite{gpt4} as an API-access LLM and Magicoder~\cite{wei2023magicoder} as an open-access LLM\footnote{These models have been shown to surpass their equivalent models of the same size in several programming tasks.}. 
When dealing with real-world code and tests, LLMs' performance can drastically degrade~\cite{pan2023understanding}. As an obvious consequence, they generate a code that does not compile, even though being asked during prompting~\cite{ouyang2023llm}. 

The ultimate goal of \approach is to repair real-world flaky tests, making it vulnerable to this limitation of LLMs. Specifically, without being compilable, passing the patch to the \validation component is worthless. As a result, \stitch sub-component of \repair attempts to resolve common compilation issues in the generated patch offline. As we will show later (\cref{subsec:RQ3}), 

% 10% = (31)/(31+268), 41% = (17)/(17+24)
\stitch contributes to 10\% and 41\% of the correct patches generated by \gpt and \magic \textit{in the first iteration} (total numbers across all iterations are 12\% and 31\%). 

These numbers are more significant, by a huge margin, than asking LLMs to resolve compilation errors through iterative textual feedback~\cite{chen2022codet,wang2023leti}. \stitch also reduces the computational cost and carbon footprint by avoiding re-promoting LLMs for fixing trivial or frequent compilation issues. 
\stitch resolves the following issues in the LLM-generated patches systematically using Algorithm~\ref{alg:stitching}:

\begin{itemize}[leftmargin=*]
    \item \textit{Inconsistency with the original code (Lines 3--7).} Patches likely differ from the original code in a few statements. Due to non-determinism intrinsic to LLMs, it is possible that the generated code, although implementing the correct repair logic, has such trivial inconsistency issues and cannot be compiled. To check for this, \stitch inspects if \textit{modifiers}, \textit{return types}, and \textit{annotations} of the test method(s) in the patch match the original code. If not, it reverts the changes at those places. In the running example of Figure~\ref{fig:overview}, the LLM-generated patch removes the \texttt{\small{public}} modifier, which prevents the test runner in JUnit4 from executing the method. Thereby, \stitch adds the \texttt{\small{public}} modifier back.

    \item \textit{Missing class dependency (Lines 10--17).} Adding new code may require importing new dependencies. If a compilation error is related to missing dependencies (i.e., \textit{missing class symbols} error), \stitch looks for the missing class in the local JDK specified in the build file of the project and imports the corresponding one that resolves the error. In the patch generated for the running example of Figure~\ref{fig:overview}, LLM replaces \texttt{\small{HashMap}} with \texttt{\small{LinkedHashMap}}, but fails to import \texttt{\small{java.util.LinkedHashMap}}. Based on the error message, \stitch looks for the class \texttt{\small{LinkedHashMap}} and adds the corresponding import to the patch. 
    %\vspace{-5pt}

    %\vspace{-5pt}
     \item \textit{Missing external dependency (Lines 18--20).} Some patches require updating the \textit{pom.xml}. For example, the patch for ID flakiness in Figure~\ref{fig:id-example} should not only import \texttt{\small{com.google.gson.Gson}} and \texttt{\small{com.google.gson.reflect.TypeToken}} to the test class, but also update \textit{pom.xml} by adding \texttt{\small{gson 2.8.6}} as a dependency (or rewrite the \texttt{\small{artifactId}} if the dependency exists).

    \item \textit{Conflicting dependencies (Lines 21--25).} LLMs may add dependencies that conflict with the existing ones. For example, adding \texttt{\small{org.junit.Assert.assertThat}} to a test that already imports \texttt{\small{org.assertj.core.api.Assertions.assertThat}} results in a compilation error due to an ambiguous reference.% to \texttt{\small{assertThat}}.
     
\end{itemize}
\vspace{-5pt}

% \vspace{-10pt}
\subsection{\validation}
\label{subsec:validation}

\approach can generate plausible patches. However, the final decision of whether the patch resolves test flakiness needs further validation. For OD-Victim tests, \validation executes the patched polluter and victim in two different orders (polluter before victim and victim before polluter) using a modified version of Surefire~\cite{surefire}.
%to execute the polluter and victim tests in varying orders. 
%This extension of Surefire allows for the precise control of the test execution order. To validate the patch, we confirm its functionality with two orders: (1) running the polluter test before the victim, and (2) running the victim test before the polluter, and 
If the victim passes in both, \approach accepts the patch as the ultimate repair. 
Given that a single polluter (\textit{P}) may pollute multiple victims (\{\textit{$V_0$, $V_1$, ... $V_m$}\}), 
%. In the process of addressing any affected victim (e.g., \textit{$V_0$}), if a patch is applied directly to the \textit{P} to clean its pollution,  
\validation also checks whether a patch removes the pollution of other victims (\{\textit{$V_1$, ... $V_m$}\}). This can reduce the need for additional re-prompts to fix each victim separately, thereby minimizing the costs.
%The flaky test suite may contain multiple victims sharing one polluter. In such cases, we first check if \approach has previously generated a correct patched polluter and validate it with the rest of the victims. If successful, we consider all the OD flakiness depending on the polluter to be resolved. 

\vspace{-5pt}
%\vspace{-8pt}
\begin{algorithm}[tbh]
%\footnotesize
% \scriptsize
\caption{ Stitching Component}
\label{alg:stitching}

\KwIn{Original Related Code $RC$, LLM-generated Code $LC$, Compilation Errors $E$}
\KwOutput{Stitched Code $SC$}
%\KwOutput{Set of helpers $H=\{H_1, \ldots, H_m\}$}

%$X, H, VB \gets \emptyset$ \\
\ForEach{$m_i \in LC$}
{
 \If{\text{hasError}($m_i$,$E$)}
    {
        $T \gets$ \text{getCorrespondingMethod($RC$,$m_i$)}  \\
        $DT \gets $identifyMethodDeclaration($T$) \\
        $DM \gets $identifyMethodDeclaration($m_i$) \\
        \If{$DT \neq DM$}
            {$SC \gets \text{revert } DM \text{ in } m_i \text{ to } DT$ \\}
            %{$SC \gets m_i$}
        $SLib \gets getJavaStandardLibs$ \\
        \ForEach{$e_i \in E$}
        {
            \If{\text{isMissingClassSymbol}$(e_i)$}
            {
                $eSymbol \gets \text{extractClassSymbol}(e_i)$ \\
                $Slib_i \gets \text{searchJavaLib}(eSymbol, SLib)$ \\
                \ForEach{$lib_r \in Slib_i$}
                {   
                    $P_r \gets \text{addImportLib}(lib_r, SC)$ \\
                    $E_r \gets \text{compile}(P_r)$ \\
                    \If{$e_i \notin E_r$}
                    {
                        $SC \gets P_r$ %\\
                        %\textbf{break}\;
                    }
                }
            }
            \If{\text{packageNotExist}$(e_i)$}
            {
                $pack \gets \text{extractMissingPackage}(e_i)$ \\
                $SC_{build} \gets \text{searchRepository}(pack)$ \\
            }
        }
        
        $PImports \gets \text{getImportStats}(m_i)$ \\
        $TImports \gets \text{getImportStats}(T)$ \\
        \ForEach{$PImp_i \in PImports$}
        {
            \If{$\text{isConflictWith}(PImp_i, TImports)$}
            {
                $SC \gets \text{exclude}(SC, PImp_i)$ \\
            }
        }
    }
}

\KwRet $SC$
\end{algorithm}
%\vspace{-8pt}
\vspace{-5pt}

For OD-Brittle, \validation executes the patched brittle test and accepts it as an ultimate fix if it passes. We did not use \idflakies to validate OD tests, mainly due to the non-determinism intrinsic to the implemented algorithm. 
%That is, \idflakies detects OD flaky tests by reordering and rerunning tests in the test suite randomly. For large test suites, which is the case in real-world projects, it cannot try all the permutations of the test execution. As a result, it may terminate without finding flakiness while it still exists. 
One threat of validating tests in isolation from other tests is \textit{overfitting}~\cite{le2018overfitting}: introducing a new bug when fixing the current one.
%Such cases happen when repairing a bug, resulting in the introduction of a new bug. In the context of this research, repairing test flakiness may result in another, new flakiness. 
While the chances of overfitting are narrow in our experiments, we performed a lightweight static analysis check to ensure the shared state between OD tests is unique to them, and no other test in the test suite has such dependency. 
The \validation uses \nondex (configured with \textit{nondexRuns=5} similar to the original paper) to validate ID patches. 
%To that end, we configured \nondex with \textit{nondexRuns=5}, the same configuration used in the original paper and several others~\cite{chen2023transforming}.  
If \nondex does not mark the patch as flaky, we accept it as the ultimate fix. 

\validation categorizes the validation outcomes into three groups: \textit{test pass}, \textit{test failure}, and \textit{compilation error}. A \textit{test pass} indicates the patch successfully resolves the issue, while the last two types indicate the patch from the current response does not fix the flakiness correctly. In such cases, the process will cycle the patch through subsequent iterations in a feedback loop for further refinement.

\subsection{\feedback}
\label{subsec:feedback}

Flaky tests are complex, and LLMs may not repair them with a single round of prompting, motivating the re-prompting of LLMs iteratively. At the end of each iteration, the \promptgeneration component takes the compilation errors or test failures as inputs, modifies the Related Code and Failure Location of the previous prompt by adding new contextual information (CI.1--CI.3), and prompts LLM again. One of the core strengths of \approach over related work that employs iterative textual feedback to improve LLM performance~\cite{wang2023leti,pan2023understanding,chen2022codet}
%~\cite{he2024iter,xia2023conversational} 
is trimming down long compilation error or test failure results (sometimes as long as $1000+$ lines) to a handful of concise contextual information (\cref{subsec:inspector}).
%\yang{Several approaches leverage iterative feedback~\cite{he2024iter,xia2023conversational} in code tasks.} Unlike other approaches that feed the entire stack trace or failure report to the model~\cite{wang2023leti,pan2023understanding,chen2022codet}, the \inspector component of \approach trims down long compilation error or test failure results (sometimes as long as $1000+$ lines) to a handful of concise contextual information (\cref{subsec:inspector}). 
This will improve the performance of \approach, as recent research shows that LLMs provide the best results when given fewer, more relevant contexts rather than larger, unfiltered ones~\cite{liu2024lost}. 
The iterative repair of a given flaky test repeats five times. However, \approach terminates the feedback loop sooner if it observes an identical compilation error in three consecutive rounds or repairs flakiness.
%of prompting or can successfully generate a valid patch. 

In-context learning~\cite{brown2020language} may improve the performance of \approach. However, including examples results in most prompts, especially those for repairing OD-Victim flakiness, exceeding LLMs' context window size. Furthermore, \approach's prompts are enriched with practical natural language examples in the Instruction section and concise context in the Related Code and Failure Location sections. Given that LLMs inherently understand instructions in natural language better than in-context examples in different modalities~\cite{wang2023aligning}, the need for in-context examples in \approach is negligible.   %Consequently, the current implementation of \approach skips in-context learning.  

\section{Evaluation}
\label{sec:evaluation}
%We evaluate the contributions of \approach under the following research questions: 
We investigate the following research questions:

\begin{enumerate}[leftmargin=*,label=\bfseries RQ\arabic*:,nosep]
    
    \item \textbf{Effectiveness in fixing Test Flakiness.}
    To what extent \approach can repair previous fixed or unfixed OD and ID flakiness in real-world projects? %What percentage of the fixes have been previously unfixed? 

    \item \textbf{Comparison with Alternative Approaches.} To what extent \approach can fix flaky tests that alternative approaches cannot? Are there flaky tests that \approach cannot fix but other techniques can?

    \item \textbf{Contribution of Different Components.} To what extent do error extraction, prompt crafting, stitching, and feedback loop help \approach to successfully repair flaky tests?

    \item \textbf{Performance.} How much does it take and cost for \approach to repair flaky tests?

\end{enumerate}

%\yang{reproducibility}

\subsection{Experimental Setup}
%In this section, we will explain our approach to selecting subject flaky tests and validating the generated plausible patches. We also explain the details of alternative approaches we used to compare the effectiveness of \approach against them in repairing flaky tests. 

%\vspace{3pt}
\textbf{Alternative Approaches.} 
Prior research focuses on repairing only one type of test flakiness; \odrepair~\cite{li2022repairing} and \ifixflakies~\cite{shi2019ifixflakies} repair Java OD tests and \ipflakies~\cite{wang2022ipflakies} repairs Python OD tests.  \dexfix~\cite{zhang2021domain} repairs ID tests. TRaF~\cite{pei2023traf} repairs a special category of NOD tests caused by asynchronous waits. We excluded TRaF from alternative approaches, as \approach currently only fixes ID and OD tests. Since most flakiness repair approaches deal with Java unit tests, we also excluded \ipflakies from the evaluation. Among the remaining tools, \dexfix is not publicly available\footnote{This was confirmed by the paper's authors.}, but their dataset is. As a result, we evaluated \approach on their dataset of ID tests without running their tool on additional ID flakiness. For OD flakiness, we compared \approach with both \odrepair and \ifixflakies (\odrepair overcomes the limitations of \ifixflakies by generating cleaner tests, while \ifixflakies can fix tests that \odrepair can not).

\textbf{Subjects.} 
%\yang{why we don't enlarge OD dataset; take odrepair dataset - sample similar numbers of ID - include dexfix dataset}
% \yang{the dataset of \odrepair consists of \hl{$327$} projects belong to \hl{$42$} projects, we excluded \hl{$28$} tests from \hl{$8$} projects which were not able to compile due to deprecated dependencies, finally, we collected \hl{$299$} OD-Victim tests from \hl{$34$} projects. }
%\reyhan{explain why our numbers (total test) are smaller than odrepair paper. Also, when discussing the results, mention the impact of removed instances}
Alternative approaches come with a dataset of ID (from \dexfix) and OD-Victim flaky tests (from \odrepair). We excluded $38$ tests from four projects in \dexfix dataset and $28$ OD-Victim tests from eight projects in \odrepair dataset that we were not able to compile or reproduce the flakiness in a reasonable amount of time, which left us with $237$ ID tests and  $299$ OD-Victim tests. %\reyhan{fill the placehodlers}
%\reyhan{why did we exclude them? please explain and fill the placeholders}. 

We further augmented these datasets with flaky tests from \idoft~\cite{idoft}, a repository of different types of flakiness in real-world projects. The reasons for augmentation are to include (1) OD-Brittle tests, which were not included in the dataset of prior work, and (2) flaky tests that were not previously fixed by human developers or automated flakiness repair techniques. 
From \idoft, 
%To collect our subjects, we use \idoft~\cite{idoft}, a repository containing more than \hl{$6387$} flaky tests of different types detected in real-world projects. 
%Among them, about \hl{$1527$} are already fixed manually by humans or automated techniques, leaving an opportunity for us to repair the remaining unfixed tests. For ID flaky tests, we started from \hl{$3642$} tests in the \idoft dataset, belonging to \hl{$476$} different projects. 
we excluded the projects that (1) were removed from the repositories mentioned in \idoft, (2) we were not able to compile with Java 8 or Java 11 due to non-trivial issues such as deprecated dependencies, (3) did not finish compilation in one hour, and (4) we were not able to reproduce their flakiness. The filtering process left us with $193$ projects with at least one module, where different modules of the same project may have different flaky tests in the \idoft dataset. Augmentation, along with the tests from the dataset of alternative approaches, provides us with 541 ID, 299 OD-Victim\footnote{No additional OD-Victim tests found in the selected projects.}, and 33 OD-Brittle tests from total $243$ projects. Among the total of $873$ tests, there are $114$, $98$, and $14$, previously unfixed ID, OD-Victim, and OD-Brittle tests.

%left us with a total o

\textbf{LLMs.} \approach is designed to work with API- and open-access LLMs. The former does not require the availability of GPU resources and is more accessible to a wider range of users. However, most API-access models, even though negligible, charge for prompting. Open-access LLMs, on the other hand, are free to use, assuming the availability of (non-trivial) GPU resources. Our experiments use \gpt~\cite{gpt4} and \magic~\cite{wei2023magicoder} as API- and open-access LLMs, given their superiority to alternative models of similar size in code synthesis~\cite{gpt4,wei2023magicoder}. LLMs are inherently non-deterministic, which impacts the reproducibility of their results. We believe this is not a threat to the validity of our results: once the synthesized code repairs the flakiness, the problem is considered to be solved. Furthermore, the iterative nature of \approach, utilizing sound program analysis as part of the approach, large-scale evaluation on real-world data (repairing 507\footnote{We count \textit{unique} tests repaired by two versions of \approach.} out of 873 flaky tests), and repairing previously unfixed flaky tests (79 previously unfixed by developers or alternative approaches) increases confidence in the rigor of the technique rather than being luck.

\subsection{RQ1: Effectiveness in Repairing Test Flakiness}
\label{subsec:RQ1}

\subsubsection{Repairing ID Flakiness}
%We were interested in measuring the extent to which \approach can repair the ID flakiness. 
%To that end, we first used \approach to repair the ID flaky tests we collected through a systematic sampling of \idoft, which contains \hl{$242$} previously fixed and \hl{$119$} previously unfixed flaky tests. These tests belong to \hl{$299$} modules across Java \hl{$214$} projects. 
%The column \textit{\approach Dataset} shows the result of this experiment. To show the results of all \hl{$214$} projects, we grouped them based on their GitHub ID. IDs associated with only a single project are grouped under the \textit{Others} category\footnote{All the detailed results of our experiments are available on the artifact website.}. 
Table~\ref{table:rq1ID} shows the result of running \approach and \dexfix on subject ID flaky tests. Columns \textit{PF} and \textit{PU} indicate the number of previously fixed and unfixed ID tests. After the automated validation, we manually checked all the repaired patches to ensure the correctness. Such false positives include deleting assert statements in the patch, surrounding them inside \texttt{\small{try/catch}} blocks, or replacing the failing assert statement with one that always passes. Column \textit{FP} shows the number of false positives identified through the manual check. The reported numbers under PF and PU do not include FP patches.
%, i.e., patches that passed the validation phase but were not semantically correct fixes.

From these results, we can see that \approach-\gpt and \approach-\magic were able to repair 
$57\%$ ($39\%$ previously unfixed) and $16\%$ ($9\%$ previously unfixed) ID flaky tests.
%in the \approach Dataset, and \magic was able to repair \hl{$19\%$} of them.} 
%Breaking down the results, \textbf{\gpt was able to repair \hl{$66\%$} of previously fixed and \hl{$50\%$} of previously unfixed ID flaky tests, \magic successfully repairs \hl{$22\%$} of previously fixed and \hl{$15\%$} of previously unfixed tests.}. 
While \magic repairs less tests compared to \gpt, it can, in fact, repair $6$ tests that \gpt cannot. As we will show in RQ3, augmenting the power of LLMs with program analysis enables some emerging abilities for smaller open-access models. 
%A pragmatic usage of \approach in fixing flaky tests may involve prompting multiple models and combining their power.  
%\reyhan{are these 7 after excluding FPs? if yes, how many of them are previously unfixed?} \yang{Yes, these 7 are excluding FPs, but all of them were previously fixed (already opened with a PR by human in iDoFT.)}
%Although the patches generated by \magic are much fewer than \gpt, there are \hl{$7$} tests that can be fixed by \magic but not \gpt. 
%We manually investigated all the patches to exclude false positives (the numbers under \approach/PF and \approach/PU columns do \textbf{not} include false positives). 
%The false positive cases we identified are mostly due to 
%\reyhan{the artifact website is not publicly accessible (or does not exist)}.

We also wanted to see to what extent \approach advances state-of-the-art ID flakiness repair technique, \dexfix.  
%To that end, we compared it with \dexfix. 
Given that \dexfix is not publicly available, to have a fair comparison, we show the performance of \approach for a subset of ID flaky tests in the last four projects that overlap with \dexfix dataset inside the parenthesis.
%\footnote{The overlap between ID tests with \dexfix dataset is only in previously fixed tests.}. 
Overall, 
%we used their dataset and evaluated the effectiveness of \approach on its flaky tests. The numbers under the \textit{\dexfix Dataset} of Table~\ref{table:rq1ID} demonstrate the results of this experiment. We do not have the notion of previously unfixed in this experiment due to the overlap of \approach Dataset and \dexfix Dataset.
 %\textbf{Overall, 
\approach-\gpt and \approach-\magic repair $52\%$ and $11\%$ of the ID flaky tests in the \dexfix Dataset,
%, and with Magicoder, it repaired \hl{$13\%$} of them, 
while \dexfix achieves $44\%$ repair success rate. The repaired ID tests by \approach-\gpt from the \dexfix dataset subsume that of \approach-\magic.
%We again investigated the successful patches for false positives and detected $4$ from each model, which had similar root causes as discussed before. We will compare the effectiveness of each technique in RQ2. 

% \yang{odrepair/ifixflakies repair OD-Brittles? if they are general to cover all categories of OD -- adding \ifixflakies for repairing od-brittles} 
% \yang{overall total numbers of each approach in figures -- added in the figure}

\subsubsection{Repairing OD Flakiness}
The numbers under \textit{OD-Victim} column of Table~\ref{table:rq1OD} compare the effectiveness of \approach with \odrepair and \ifixflakies. \textit{OD-Brittle} column only compares \approach and \ifixflakies, as \odrepair cannot fix such flakiness without knowing corresponding state-unsetters (a test that pollutes the state for brittle tests)~\cite{chen2023transforming}. 
%\reyhan{why \odrepair cannot fix them?} 
%\yang{\odrepair is specifically designed to generate patches for victims where the polluter is known. Therefore, its ability to repair OD-Brittles is limited if only receiving OD-Brittles as input.} 

Similar to the previous experiment, we manually checked and excluded false positives from the results.
%These results show that \textbf{
\approach-\gpt can repair $58\%$ ($27\%$ previously unfixed) OD-Victim tests. \approach-\magic repairs $27\%$, all subsumed by \approach-\gpt. 
%\reyhan{similar to OD-Brittle and ID, report the percentage of previously unfixed numbers (for gpt)}.
%, \magic was able to achieve \hl{$27\%$}, while 
On the other hand, \odrepair and \ifixflakies repair $45\%$ and $40\%$ of OD-Victim tests. To recall, we have to exclude $28$ tests from \odrepair and $38$ tests from \dexfix dataset due to non-trivial deprecated dependencies or non-reproducible flakiness, out of which, \odrepair successfully repairs only five, and \dexfix successfully repairs $15$. This still makes \approach superior to \odrepair and \dexfix, given the notable gap in repairing OD-Victim and ID tests.
%tests we have to exclude from \odrepair dataset, $5$ tests that were successfully fixed, should not threaten the validity of the findings. This is because \approach already repairs $37$ more tests than \odrepair even with the omission. 
Fixing OD-Brittle is a tie-in competition for \approach-\gpt and \ifixflakies, \approach achieves $51\%$ success rate and \ifixflakies achieves $39\%$.
%both achieving \hl{$51\%$} success rate. However, only around half of the repaired tests by these two techniques overlap (\cref{subsec:RQ2}).
\approach-\magic can only repair $9\%$ of the OD-Brittle flaky tests, where one of them could not be fixed by \approach-\gpt. 
%Regarding OD-Brittle tests, \approach-\gpt can successfully repair \hl{$51\%$} (\hl{$20\%$} previously unfixed), while \approach-\magic can repair \hl{$9\%$} of them, \ifixflakies achieves a successful repair rate \hl{$51\%$}. The repaired OD-Victim tests by \approach-\gpt from the \dexfix dataset subsume that of \approach-\magic, while there is one OD-Brittle test that can be fixed by \approach-\magic but not \approach-\gpt.

\textbf{As demonstrated, \approach was able to repair $79$ previously unfixed ID and OD flaky tests. We have opened PRs for such fixes, where $19$ of them have been accepted and merged by the time of submission.
%\footnote{The links to the opened PRs are available on the artifact website~\cite{website}.}.
We consider this ability of \approach to make it superior to flakiness repair approaches in particular, and to a wider range of LLM-based program repair techniques, in general. Comparing \approach with general State-of-the-art APR techniques~\cite{xia2023conversational,he2024iter}, even those that leverage LLMs such as \gpt~\cite{xia2023keep}, have been only proven to be effective on known datasets such as Defects4J~\cite{defects4j} and QuixBugs~\cite{quix}. In contrast, \approach repairs many flaky tests from real-world projects, where humans or automated techniques previously could not repair a reasonable number. As we will show in subsequent research questions, this power comes from the synergy of LLMs and symbolic approaches, not just LLMs.}

%\yang{Comparing \approach with flakiness repair tools, we do not rely on existing cleaner/state-setter tests to generate patches, and we can also repair general types of flaky tests. 
% \yang{comparison v.s. 1)flakiness repair tools, 1-LLMs; 2-fix general types; 3-fix more tests; 2)general APR tools}

\begin{table}[t]
    \setlength{\tabcolsep}{0.1pt}
    \centering
    % \footnotesize
    \scriptsize
    % \tiny 
    % \small
    \vspace{15pt}
    \caption{\small Effectiveness of \approach and \dexfix in repairing ID flakiness. \textbf{P}: Projects; \textbf{M}: Modules; \textbf{PF}: Previously Fixed; \textbf{PU}: Previously Unfixed; \textbf{FP}: False Positive. \colorbox{patchgreen}{Green} rows indicate the superiority of \approach, and the \colorbox{maroon!30}{red} row indicates the superiority of \dexfix. The white rows belong to augmented tests. 
    %\reyhan{(1) remove the \approach Dataset row from the top of table (2) under \#Tests/PF, please include the number of \dexfix tests inside parenthesis} %\reyhan{the margin from top is very small. fix it please} \yang{trying to figure out the root cause}
    %(GCP: GoogleCloudPlatform; SP: spring-projects.)
    %\reyhan{ (1) merge all IDs into one dataset (2) remove the "\dexfix Dataset" super column and move "\dexfix fixed" numbers under the "\approach Dataset" (3) actually, no need for a super column dataset, as all will be ID tests}
    }
    \vspace{-5pt}
    \renewcommand{\arraystretch}{0.94}
\begin{tabular}{|c|c|c|wc{2em}||wc{4em}|wc{2em}|wc{2em}||wc{4em}|wc{2em}|wc{2em}||c|}
\hline
\multicolumn{1}{|c|}{\multirow{2}{*}{GitHub ID}} &
\multicolumn{1}{c|}{\multirow{2}{*}{\#P [\#M]}} &
\multicolumn{2}{c|}{\#Tests} &
\multicolumn{3}{c|}{GPT-4}& 
\multicolumn{3}{c|}{Magicoder}& 
% \multicolumn{3}{c|}{Code Llama}& 
% \multicolumn{1}{c|}{\multirow{2}{*}{\# ID Tests}} &
% \multicolumn{1}{c|}{\#Tests} &
\multicolumn{1}{c|}{\dexfix}  \\
% & \multicolumn{6}{c|}{\dexfix Dataset} \\

\multicolumn{1}{|c|}{} &
\multicolumn{1}{c|}{} &
% \multicolumn{1}{c|}{} & 
\multicolumn{1}{c|}{PF} &
\multicolumn{1}{c|}{PU} &
\multicolumn{1}{c|}{PF} &
\multicolumn{1}{c|}{PU} &
\multicolumn{1}{c|}{FP} &
\multicolumn{1}{c|}{PF} &
\multicolumn{1}{c|}{PU} &
\multicolumn{1}{c|}{FP} &
\multicolumn{1}{c|}{\#Fixed} \\ \hline

FasterXML & 10 [9] & 12  & 2 & {\bf9}  & {\bf2} & 1 & 1  &0 & 1 & -  \\
SAP & 4 [2] & 4  & 1 & {\bf2}  & 0 & 0 & 1  &0 & 0 & -  \\
IBM & 4 [4] & 4  & 0 & {\bf3}  & 0 & 0 & 0  &0 & 0 & -  \\
adobe & 3 [2] & 3  & 1 & {\bf2}  & 0 & 0 & 1  &0 & 0 & -  \\
DataDog & 3 [1] & 2  & 1 & {\bf2}  & {\bf1} & 0 & 0  &0 & 1 & -  \\
oracle & 3 [3] & 3  & 0 & {\bf1}  & 0 & 0 & {\bf1}  &0 & 0 & -  \\
wildfly & 2 [5] & 3  & 2 & {\bf1}  & 0 & 0 & {\bf1}  &0 & 1 & -  \\
intel & 2 [1] & 1  & 1 & {\bf1}  & 0 & 0 & {\bf1}  &0 & 0 & -  \\
networknt & 2 [5] & 5  & 0 & {\bf3}  & 0 & 0 & 2  &0 & 0 & -  \\
gchq & 2 [5] & 4  & 2 & {\bf4}  & 0 & 1 & 1  &0 & 0 & -  \\
opengoofy & 2 [3] & 4  & 0 & {\bf4}  & 0 & 1 & 2  &0 & 0 & -  \\
eclipse-ee4j & 2 [6] & 1  & 5 & 0  & {\bf3} & 0 & 0  &1 & 0 & -  \\
SP~\tablefootnote{spring-projects} & 2 [2] & 2  & 0 & {\bf2}  & 0 & 0 & 0  &0 & 0 & -  \\
HubSpot & 2 [2] & 2  & 0 & {\bf2}  & 0 & 0 & 1  &0 & 0 & -  \\
twitter & 2 [2] & 2  & 0 & {\bf1}  & 0 & 0 & 0  &0 & 0 & -  \\
dromara & 2 [3] & 3  & 0 & {\bf2}  & 0 & 0 & 0  &0 & 0 & -  \\
GCP~\tablefootnote{GoogleCloudPlatform} & 2 [1] & 1  & 1 & {\bf1}  & 0 & 0 & {\bf1}  &0 & 0 & -  \\
eBay & 2 [2] & 1  & 1 & 0  & {\bf1} & 0 & 0  &0 & 0 & -  \\
jdereg & 2 [1] & 0  & 2 & 0  & 0 & 0 & 0  &0 & 1 & -  \\
 
 apache & 36 [87] & 
\cellcolor{patchgreen}{136 (98)} & \cellcolor{patchgreen}{46} & \cellcolor{patchgreen}{{\bf75 (49)}} & \cellcolor{patchgreen}{16} & \cellcolor{patchgreen}{0} & \cellcolor{patchgreen}{17 (5)} &\cellcolor{patchgreen}{0} & \cellcolor{patchgreen}{9} & \cellcolor{patchgreen}{29}  \\
 
square & 4 [4] &
\cellcolor{patchgreen}{3 (9)} & \cellcolor{patchgreen}{9} & \cellcolor{patchgreen}{{\bf2 (1)}} & \cellcolor{patchgreen}{2} & \cellcolor{patchgreen}{0} & \cellcolor{patchgreen}{0 (0)} &\cellcolor{patchgreen}{0} & \cellcolor{patchgreen}{0} & \cellcolor{patchgreen}{0}  \\

 AC~\tablefootnote{apolloconfig} & 2 [5] &
\cellcolor{patchgreen}{9 (4)} & \cellcolor{patchgreen}{0} & \cellcolor{patchgreen}{{\bf7 (2)}} & \cellcolor{patchgreen}{0} & \cellcolor{patchgreen}{0} & \cellcolor{patchgreen}{4 (1)} &\cellcolor{patchgreen}{0} & \cellcolor{patchgreen}{0} & \cellcolor{patchgreen}{0}  \\

intuit & 2 [2] &
\cellcolor{patchgreen}{1 (2)} & \cellcolor{patchgreen}{2} & \cellcolor{patchgreen}{{\bf1 (1)}} & \cellcolor{patchgreen}{1} & \cellcolor{patchgreen}{0} & \cellcolor{patchgreen}{0 (0)} &\cellcolor{patchgreen}{0} & \cellcolor{patchgreen}{0} & \cellcolor{patchgreen}{0}  \\

 Others & 114 [143] &
\cellcolor{patchgreen}{182 (91)} & \cellcolor{patchgreen}{37} & \cellcolor{patchgreen}{{\bf117 (51)}} & \cellcolor{patchgreen}{17} & \cellcolor{patchgreen}{3} & \cellcolor{patchgreen}{40 (18)} &\cellcolor{patchgreen}{9} & \cellcolor{patchgreen}{5} & \cellcolor{patchgreen}{50}  \\

 alibaba & 4 [6] &
\cellcolor{maroon!30}{39 (33)} & \cellcolor{maroon!30}{1} & \cellcolor{maroon!30}{25 (20) }& \cellcolor{maroon!30}{ 1} & \cellcolor{maroon!30}{0} & \cellcolor{maroon!30}{3 (2) }&\cellcolor{maroon!30}{0} & \cellcolor{maroon!30}{0} & \cellcolor{maroon!30}{ {\bf25}}  \\
\hline
\rowcolor{mygray} Total & 215 [306] & 427 (237) & 114 & {\bf267 (124)} & 44 & 6 & 77 (26) &10 & 18 & 104 \\

\hline
\end{tabular}
    \label{table:rq1ID}
    \vspace{-15pt}
\end{table}

\begin{table*}[t]
    \setlength{\tabcolsep}{0.5pt}
    \centering
    % \footnotesize
    \small
    \caption{ Effectiveness of \approach, \odrepair and \ifixflakies in repairing OD flakiness. \textbf{P}: Github Projects; \textbf{M}: Modules; \textbf{PF}: Previously Fixed ID tests; \textbf{PU}: Previously Unfixed ID tests; \textbf{FP}: False Positive. \colorbox{patchgreen}{Green} rows indicate the superiority of \approach, \colorbox{myyellow}{yellow} rows indicate tie, \colorbox{maroon!30}{red} rows indicate the superiority of alternative approaches, and the white row indicates cases where none of the techniques repaired flaky tests.   
    %Green: \approach fixes max tests; Yellow: \approach fixes same number of tests with others; Red: \approach fixes no tests.
    %\reyhan{Under column \#P [\#M], please replace () with []}
    %\reyhan{to be consistent with Table 1, please merge projects and module columns}
    %\reyhan{In this table, distinguish the max values among the models with bold font (I have already done a few rows). we should use some colors for better presentation. highlight the rows where we find more flakiness with green, ties with yellow, and red where we cannot find any (don't change font color, but table row background). there are also some rows where no tool can fix the flakiness. Move them to the end of od-victim projects. Use the same strategy in Table 1}
    }
    %\vspace{-5pt}
    \renewcommand{\arraystretch}{0.94}
\begin{tabular}{|c|c|wc{2em}|wc{2em}||wc{2em}|wc{2em}|wc{2em}||wc{2em}|wc{2em}|wc{2em}||c||c|wc{2em}|wc{2em}||wc{2em}|wc{2em}|wc{2em}||wc{2em}|wc{2em}|wc{2em}||c|}

\hline
\multicolumn{1}{|c|}{\multirow{3}{*}{GitHub ID}} &
\multirow{3}{*}{ \#P [\#M]} &
% \multirow{3}{*}{ \#Modules} &
\multicolumn{10}{c|}{OD-Victim}& 
\multicolumn{9}{c|}{OD-Brittle} \\
&                             
% & 
&
\multicolumn{2}{c||}{\#Tests} &
\multicolumn{3}{c||}{ \gpt } &
\multicolumn{3}{c||}{ \magic } &
\multirow{2}{*}{\odrepair} &
\multirow{2}{*}{\ifixflakies} &
\multicolumn{2}{c||}{\#Tests} &
\multicolumn{3}{c||}{ \gpt } &
\multicolumn{3}{c||}{ \magic } &
\multirow{2}{*}{\ifixflakies}
\\

& & PF&  PU & PF& PU & FP& PF& PU & FP& & & PF  &  PU &  PF &  PU &  FP & PF &  PU &  FP & \\
\hline

winder &  1 [2] & 
\cellcolor{patchgreen}{5}&\cellcolor{patchgreen}{1}&\cellcolor{patchgreen}{{\bf5}}&\cellcolor{patchgreen}{{\bf1}}&\cellcolor{patchgreen}{0}&\cellcolor{patchgreen}{0}&\cellcolor{patchgreen}{0}&\cellcolor{patchgreen}{0}&\cellcolor{patchgreen}{5}&\cellcolor{patchgreen}{5}&\cellcolor{patchgreen}{-}&\cellcolor{patchgreen}{-}&\cellcolor{patchgreen}{-}&\cellcolor{patchgreen}{-}&\cellcolor{patchgreen}{-}&\cellcolor{patchgreen}{-}&\cellcolor{patchgreen}{-}&\cellcolor{patchgreen}{-}&\cellcolor{patchgreen}{-}\\

tbsalling & 1 [1] & 
\cellcolor{patchgreen}{2}&\cellcolor{patchgreen}{0}&\cellcolor{patchgreen}{{\bf2}}&\cellcolor{patchgreen}{0}&\cellcolor{patchgreen}{0}&\cellcolor{patchgreen}{0}&\cellcolor{patchgreen}{0}&\cellcolor{patchgreen}{0}&\cellcolor{patchgreen}{0}&\cellcolor{patchgreen}{0}&\cellcolor{patchgreen}{-}&\cellcolor{patchgreen}{-}&\cellcolor{patchgreen}{-}&\cellcolor{patchgreen}{-}&\cellcolor{patchgreen}{-}&\cellcolor{patchgreen}{-}&\cellcolor{patchgreen}{-}&\cellcolor{patchgreen}{-}&\cellcolor{patchgreen}{-}\\

tools4j & 1 [1] & 
\cellcolor{patchgreen}{1}&\cellcolor{patchgreen}{0}&\cellcolor{patchgreen}{{\bf1}}&\cellcolor{patchgreen}{0}&\cellcolor{patchgreen}{0}&\cellcolor{patchgreen}{0}&\cellcolor{patchgreen}{0}&\cellcolor{patchgreen}{0}&\cellcolor{patchgreen}{0}&\cellcolor{patchgreen}{0}&\cellcolor{patchgreen}{-}&\cellcolor{patchgreen}{-}&\cellcolor{patchgreen}{-}&\cellcolor{patchgreen}{-}&\cellcolor{patchgreen}{-}&\cellcolor{patchgreen}{-}&\cellcolor{patchgreen}{-}&\cellcolor{patchgreen}{-}&\cellcolor{patchgreen}{-}\\

yangfuhai & 1 [1] & 
\cellcolor{patchgreen}{0}&\cellcolor{patchgreen}{6}&\cellcolor{patchgreen}{0}&\cellcolor{patchgreen}{{\bf6}}&\cellcolor{patchgreen}{0}&\cellcolor{patchgreen}{0}&\cellcolor{patchgreen}{0}&\cellcolor{patchgreen}{0}&\cellcolor{patchgreen}{0}&\cellcolor{patchgreen}{0}&\cellcolor{patchgreen}{-}&\cellcolor{patchgreen}{-}&\cellcolor{patchgreen}{-}&\cellcolor{patchgreen}{-}&\cellcolor{patchgreen}{-}&\cellcolor{patchgreen}{-}&\cellcolor{patchgreen}{-}&\cellcolor{patchgreen}{-}&\cellcolor{patchgreen}{-}\\

jnr & 1 [1] & 
\cellcolor{patchgreen}{0}&\cellcolor{patchgreen}{4}&\cellcolor{patchgreen}{0}&\cellcolor{patchgreen}{{\bf3}}&\cellcolor{patchgreen}{0}&\cellcolor{patchgreen}{0}&\cellcolor{patchgreen}{0}&\cellcolor{patchgreen}{0}&\cellcolor{patchgreen}{0}&\cellcolor{patchgreen}{0}&\cellcolor{patchgreen}{-}&\cellcolor{patchgreen}{-}&\cellcolor{patchgreen}{-}&\cellcolor{patchgreen}{-}&\cellcolor{patchgreen}{-}&\cellcolor{patchgreen}{-}&\cellcolor{patchgreen}{-}&\cellcolor{patchgreen}{-}&\cellcolor{patchgreen}{-}\\

Activiti & 1 [1] & 
\cellcolor{patchgreen}{0}&\cellcolor{patchgreen}{11}&\cellcolor{patchgreen}{0}&\cellcolor{patchgreen}{{\bf9}}&\cellcolor{patchgreen}{0}&\cellcolor{patchgreen}{0}&\cellcolor{patchgreen}{0}&\cellcolor{patchgreen}{0}&\cellcolor{patchgreen}{0}&\cellcolor{patchgreen}{0}&\cellcolor{patchgreen}{0}&\cellcolor{patchgreen}{8}&\cellcolor{patchgreen}{0}&\cellcolor{patchgreen}{{\bf7}}&\cellcolor{patchgreen}{0}&\cellcolor{patchgreen}{0}&\cellcolor{patchgreen}{0}&\cellcolor{patchgreen}{0}&\cellcolor{patchgreen}{0}\\

wildfly & 1 [1] & 
\cellcolor{patchgreen}{37}&\cellcolor{patchgreen}{0}&\cellcolor{patchgreen}{{\bf37}}&\cellcolor{patchgreen}{0}&\cellcolor{patchgreen}{0}&\cellcolor{patchgreen}{{\bf37}}&\cellcolor{patchgreen}{0}&\cellcolor{patchgreen}{0}&\cellcolor{patchgreen}{0}&\cellcolor{patchgreen}{0}&\cellcolor{patchgreen}{1}&\cellcolor{patchgreen}{0}&\cellcolor{patchgreen}{0}&\cellcolor{patchgreen}{0}&\cellcolor{patchgreen}{1}&\cellcolor{patchgreen}{0}&\cellcolor{patchgreen}{0}&\cellcolor{patchgreen}{0}&\cellcolor{patchgreen}{{\bf1}}\\
 
vmware & 1 [1] & 
\cellcolor{patchgreen}{1}&\cellcolor{patchgreen}{0}&\cellcolor{patchgreen}{0}&\cellcolor{patchgreen}{0}&\cellcolor{patchgreen}{0}&\cellcolor{patchgreen}{0}&\cellcolor{patchgreen}{0}&\cellcolor{patchgreen}{0}&\cellcolor{patchgreen}{0}&\cellcolor{patchgreen}{0}&\cellcolor{patchgreen}{0}&\cellcolor{patchgreen}{1}&\cellcolor{patchgreen}{0}&\cellcolor{patchgreen}{{\bf1}}&\cellcolor{patchgreen}{0}&\cellcolor{patchgreen}{0}&\cellcolor{patchgreen}{0}&\cellcolor{patchgreen}{0}&\cellcolor{patchgreen}{0}\\

wikidata & 1 [1] & 
\cellcolor{patchgreen}{2}&\cellcolor{patchgreen}{0}&\cellcolor{patchgreen}{{\bf2}}&\cellcolor{patchgreen}{0}&\cellcolor{patchgreen}{0}&\cellcolor{patchgreen}{0}&\cellcolor{patchgreen}{0}&\cellcolor{patchgreen}{0}&\cellcolor{patchgreen}{0}&\cellcolor{patchgreen}{{\bf2}}&\cellcolor{patchgreen}{3}&\cellcolor{patchgreen}{0}&\cellcolor{patchgreen}{{\bf3}}&\cellcolor{patchgreen}{0}&\cellcolor{patchgreen}{0}&\cellcolor{patchgreen}{2}&\cellcolor{patchgreen}{0}&\cellcolor{patchgreen}{0}&\cellcolor{patchgreen}{{\bf3}}\\

alibaba & 2 [2] & \cellcolor{patchgreen}{4} & \cellcolor{patchgreen}{0} & \cellcolor{patchgreen}{{\bf3}} & \cellcolor{patchgreen}{0} & \cellcolor{patchgreen}{1} & \cellcolor{patchgreen}{1} & \cellcolor{patchgreen}{0} & \cellcolor{patchgreen}{0} & \cellcolor{patchgreen}{1} & \cellcolor{patchgreen}{1} & 
\cellcolor{maroon!30}{1} & \cellcolor{maroon!30}{1} & \cellcolor{maroon!30}{0} & \cellcolor{maroon!30}{1} & \cellcolor{maroon!30}{1} & \cellcolor{maroon!30}{1} & \cellcolor{maroon!30}{0} & \cellcolor{maroon!30}{1} & \cellcolor{maroon!30}{{\bf2}} \\ 
apache & 8 [21] & \cellcolor{maroon!30}{45} & \cellcolor{maroon!30}{41} & \cellcolor{maroon!30}{6} & \cellcolor{maroon!30}{7} & \cellcolor{maroon!30}{2} & \cellcolor{maroon!30}{0} & \cellcolor{maroon!30}{0} & \cellcolor{maroon!30}{0} & \cellcolor{maroon!30}{{\bf33}} 
 &  \cellcolor{maroon!30}{10} &  \cellcolor{patchgreen}{5} & \cellcolor{patchgreen}{1} & \cellcolor{patchgreen}{{\bf2}} & \cellcolor{patchgreen}{0} & \cellcolor{patchgreen}{0} & \cellcolor{patchgreen}{0} & \cellcolor{patchgreen}{0} & \cellcolor{patchgreen}{0} & \cellcolor{patchgreen}{1} \\
 
fhoeben & 1 [1] &
\cellcolor{myyellow}{-} &\cellcolor{myyellow}{-} &\cellcolor{myyellow}{-} &\cellcolor{myyellow}{-} &\cellcolor{myyellow}{-} &\cellcolor{myyellow}{-} &\cellcolor{myyellow}{-} &\cellcolor{myyellow}{-} &\cellcolor{myyellow}{-} &\cellcolor{myyellow}{-} &\cellcolor{myyellow}{1} &\cellcolor{myyellow}{0} &\cellcolor{myyellow}{{\bf1}} &\cellcolor{myyellow}{0} &\cellcolor{myyellow}{0} &\cellcolor{myyellow}{0} &\cellcolor{myyellow}{0} &\cellcolor{myyellow}{1} &\cellcolor{myyellow}{{\bf1}}\\

OpenHFT & 1 [1] &
\cellcolor{myyellow}{-} &\cellcolor{myyellow}{-} &\cellcolor{myyellow}{-} &\cellcolor{myyellow}{-} &\cellcolor{myyellow}{-} &\cellcolor{myyellow}{-} &\cellcolor{myyellow}{-} &\cellcolor{myyellow}{-} &\cellcolor{myyellow}{-} &\cellcolor{myyellow}{-} &\cellcolor{myyellow}{2} &\cellcolor{myyellow}{0} &\cellcolor{myyellow}{{\bf2}} &\cellcolor{myyellow}{0} &\cellcolor{myyellow}{0} &\cellcolor{myyellow}{0} &\cellcolor{myyellow}{0} &\cellcolor{myyellow}{0} &\cellcolor{myyellow}{{\bf2}}\\

undertow-io & 1 [1] &
\cellcolor{myyellow}{1} &\cellcolor{myyellow}{0} &\cellcolor{myyellow}{{\bf1}} &\cellcolor{myyellow}{0} &\cellcolor{myyellow}{0} &\cellcolor{myyellow}{0} &\cellcolor{myyellow}{0} &\cellcolor{myyellow}{0} &\cellcolor{myyellow}{{\bf1}} &\cellcolor{myyellow}{{\bf1}} &\cellcolor{myyellow}{-} &\cellcolor{myyellow}{-} &\cellcolor{myyellow}{-} &\cellcolor{myyellow}{-} &\cellcolor{myyellow}{-} &\cellcolor{myyellow}{-} &\cellcolor{myyellow}{-} &\cellcolor{myyellow}{-} &\cellcolor{myyellow}{-}\\

 kevinsawicki & 1 [1] &
\cellcolor{myyellow}{28} &\cellcolor{myyellow}{0} &\cellcolor{myyellow}{{\bf28}} &\cellcolor{myyellow}{0} &\cellcolor{myyellow}{0} &\cellcolor{myyellow}{0} &\cellcolor{myyellow}{0} &\cellcolor{myyellow}{0} &\cellcolor{myyellow}{{\bf28}} &\cellcolor{myyellow}{{\bf28}} &\cellcolor{myyellow}{-} &\cellcolor{myyellow}{-} &\cellcolor{myyellow}{-} &\cellcolor{myyellow}{-} &\cellcolor{myyellow}{-} &\cellcolor{myyellow}{-} &\cellcolor{myyellow}{-} &\cellcolor{myyellow}{-} &\cellcolor{myyellow}{-}\\

 Thomas-S-B & 1 [1] &
\cellcolor{myyellow}{46} &\cellcolor{myyellow}{0} &\cellcolor{myyellow}{{\bf46}} &\cellcolor{myyellow}{0} &\cellcolor{myyellow}{0} &\cellcolor{myyellow}{40} &\cellcolor{myyellow}{0} &\cellcolor{myyellow}{0} &\cellcolor{myyellow}{{\bf46}} &\cellcolor{myyellow}{{\bf46}} &\cellcolor{myyellow}{-} &\cellcolor{myyellow}{-} &\cellcolor{myyellow}{-} &\cellcolor{myyellow}{-} &\cellcolor{myyellow}{-} &\cellcolor{myyellow}{-} &\cellcolor{myyellow}{-} &\cellcolor{myyellow}{-} &\cellcolor{myyellow}{-}\\

 ktuukkan & 1 [1] &
\cellcolor{myyellow}{12} &\cellcolor{myyellow}{0} &\cellcolor{myyellow}{{\bf12}} &\cellcolor{myyellow}{0} &\cellcolor{myyellow}{0} &\cellcolor{myyellow}{0} &\cellcolor{myyellow}{0} &\cellcolor{myyellow}{0} &\cellcolor{myyellow}{{\bf12}} &\cellcolor{myyellow}{{\bf12}} &\cellcolor{myyellow}{-} &\cellcolor{myyellow}{-} &\cellcolor{myyellow}{-} &\cellcolor{myyellow}{-} &\cellcolor{myyellow}{-} &\cellcolor{myyellow}{-} &\cellcolor{myyellow}{-} &\cellcolor{myyellow}{-} &\cellcolor{myyellow}{-}\\

 google & 1 [1] &
\cellcolor{myyellow}{1} &\cellcolor{myyellow}{0} &\cellcolor{myyellow}{{\bf1}} &\cellcolor{myyellow}{0} &\cellcolor{myyellow}{0} &\cellcolor{myyellow}{0} &\cellcolor{myyellow}{0} &\cellcolor{myyellow}{0} &\cellcolor{myyellow}{{\bf1}} &\cellcolor{myyellow}{0} &\cellcolor{myyellow}{-} &\cellcolor{myyellow}{-} &\cellcolor{myyellow}{-} &\cellcolor{myyellow}{-} &\cellcolor{myyellow}{-} &\cellcolor{myyellow}{-} &\cellcolor{myyellow}{-} &\cellcolor{myyellow}{-} &\cellcolor{myyellow}{-}\\
 
spring-projects & 2 [3] & \cellcolor{myyellow}{2} & \cellcolor{myyellow}{11} & \cellcolor{myyellow}{{\bf2}} & \cellcolor{myyellow}{0} & \cellcolor{myyellow}{0} & \cellcolor{myyellow}{{\bf2}} & \cellcolor{myyellow}{0} & \cellcolor{myyellow}{1} & \cellcolor{myyellow}{0} & \cellcolor{myyellow}{{\bf2}} & 
0 & 1 & 0 & 0 & 0 & 0 & 0 & 0 & 0 \\ 

 ConsenSys & 1 [2] &
\cellcolor{maroon!30}{3}&\cellcolor{maroon!30}{3}&\cellcolor{maroon!30}{0}&\cellcolor{maroon!30}{0}&\cellcolor{maroon!30}{0}&\cellcolor{maroon!30}{0}&\cellcolor{maroon!30}{0}&\cellcolor{maroon!30}{0}&\cellcolor{maroon!30}{0}&\cellcolor{maroon!30}{{\bf3}}&\cellcolor{maroon!30}{-}&\cellcolor{maroon!30}{-}&\cellcolor{maroon!30}{-}&\cellcolor{maroon!30}{-}&\cellcolor{maroon!30}{-}&\cellcolor{maroon!30}{-}&\cellcolor{maroon!30}{-}&\cellcolor{maroon!30}{-}&\cellcolor{maroon!30}{-}\\

 ctco & 1 [1] &
\cellcolor{maroon!30}{1}&\cellcolor{maroon!30}{0}&\cellcolor{maroon!30}{0}&\cellcolor{maroon!30}{0}&\cellcolor{maroon!30}{0}&\cellcolor{maroon!30}{0}&\cellcolor{maroon!30}{0}&\cellcolor{maroon!30}{0}&\cellcolor{maroon!30}{0}&\cellcolor{maroon!30}{{\bf1}}&\cellcolor{maroon!30}{-}&\cellcolor{maroon!30}{-}&\cellcolor{maroon!30}{-}&\cellcolor{maroon!30}{-}&\cellcolor{maroon!30}{-}&\cellcolor{maroon!30}{-}&\cellcolor{maroon!30}{-}&\cellcolor{maroon!30}{-}&\cellcolor{maroon!30}{-}\\

 dropwizard & 1 [1] &
\cellcolor{maroon!30}{1}&\cellcolor{maroon!30}{0}&\cellcolor{maroon!30}{0}&\cellcolor{maroon!30}{0}&\cellcolor{maroon!30}{0}&\cellcolor{maroon!30}{0}&\cellcolor{maroon!30}{0}&\cellcolor{maroon!30}{0}&\cellcolor{maroon!30}{0}&\cellcolor{maroon!30}{{\bf1}}&\cellcolor{maroon!30}{-}&\cellcolor{maroon!30}{-}&\cellcolor{maroon!30}{-}&\cellcolor{maroon!30}{-}&\cellcolor{maroon!30}{-}&\cellcolor{maroon!30}{-}&\cellcolor{maroon!30}{-}&\cellcolor{maroon!30}{-}&\cellcolor{maroon!30}{-}\\

 networknt & 1 [6] &
\cellcolor{maroon!30}{9}&\cellcolor{maroon!30}{5}&\cellcolor{maroon!30}{0}&\cellcolor{maroon!30}{0}&\cellcolor{maroon!30}{0}&\cellcolor{maroon!30}{0}&\cellcolor{maroon!30}{0}&\cellcolor{maroon!30}{0}&\cellcolor{maroon!30}{{\bf9}}&\cellcolor{maroon!30}{{\bf9}}&\cellcolor{maroon!30}{-}&\cellcolor{maroon!30}{-}&\cellcolor{maroon!30}{-}&\cellcolor{maroon!30}{-}&\cellcolor{maroon!30}{-}&\cellcolor{maroon!30}{-}&\cellcolor{maroon!30}{-}&\cellcolor{maroon!30}{-}&\cellcolor{maroon!30}{-}\\

 hexagonframework & 1 [1] &
\cellcolor{maroon!30}{-}&\cellcolor{maroon!30}{-}&\cellcolor{maroon!30}{-}&\cellcolor{maroon!30}{-}&\cellcolor{maroon!30}{-}&\cellcolor{maroon!30}{-}&\cellcolor{maroon!30}{-}&\cellcolor{maroon!30}{-}&\cellcolor{maroon!30}{-}&\cellcolor{maroon!30}{-}&\cellcolor{maroon!30}{1}&\cellcolor{maroon!30}{0}&\cellcolor{maroon!30}{0}&\cellcolor{maroon!30}{0}&\cellcolor{maroon!30}{0}&\cellcolor{maroon!30}{0}&\cellcolor{maroon!30}{0}&\cellcolor{maroon!30}{0}&\cellcolor{maroon!30}{{\bf1}}\\

 pinterest & 1 [1] &
\cellcolor{maroon!30}{-}&\cellcolor{maroon!30}{-}&\cellcolor{maroon!30}{-}&\cellcolor{maroon!30}{-}&\cellcolor{maroon!30}{-}&\cellcolor{maroon!30}{-}&\cellcolor{maroon!30}{-}&\cellcolor{maroon!30}{-}&\cellcolor{maroon!30}{-}&\cellcolor{maroon!30}{-}&\cellcolor{maroon!30}{2}&\cellcolor{maroon!30}{0}&\cellcolor{maroon!30}{0}&\cellcolor{maroon!30}{0}&\cellcolor{maroon!30}{0}&\cellcolor{maroon!30}{0}&\cellcolor{maroon!30}{0}&\cellcolor{maroon!30}{0}&\cellcolor{maroon!30}{{\bf2}}\\
%\rowcolor{maroon!30} 
%%%vaadin & 1 [1] & 0 & 1 & 0 & 0 & 0 & 0 & 0 & 0 & 0 & 0 & - & - & - & - & - & - & - & - & - \\ 
%\rowcolor{maroon!30} 
%%%danfickle & 1 [1] & 0 & 1 & 0 & 0 & 0 & 0 & 0 & 0 & 0 & 0 & - & - & - & - & - & - & - & - & - \\ 
%\rowcolor{maroon!30} 
%%%jenkinsci & 1 [1] & 0 & 9 & 0 & 0 & 0 & 0 & 0 & 0 & 0 & 0 & - & - & - & - & - & - & - & - & - \\ 
%\rowcolor{maroon!30} 
%%%c2mon & 1 [1] & 0 & 1 & 0 & 0 & 0 & 0 & 0 & 0 & 0 & 0 & - & - & - & - & - & - & - & - & - \\ 
%\rowcolor{maroon!30} 
%%%CloudSlang & 1 [1] & 0 & 3 & 0 & 0 & 0 & 0 & 0 & 0 & 0 & 0 & - & - & - & - & - & - & - & - & - \\ 
%\rowcolor{maroon!30} 
%%%jitsi & 1 [1] & 0 & 1 & 0 & 0 & 0 & 0 & 0 & 0 & 0 & 0 & - & - & - & - & - & - & - & - & - \\ 
%\rowcolor{maroon!30} 
%%%flaxsearch & 1 [1] & - & - & - & - & - & - & - & - & - & - & 2 & 0 & 0 & 0 & 0 & 0 & 0 & 0 & 0 \\ 
%\rowcolor{maroon!30} 
%%%javadelight & 1 [1] & - & - & - & - & - & - & - & - & - & - & 0 & 1 & 0 & 0 & 0 & 0 & 0 & 0 & 0 \\ 
%\rowcolor{maroon!30} 
%%%querydsl & 1 [1] & - & - & - & - & - & - & - & - & - & - & 1 & 1 & 0 & 0 & 0 & 0 & 0 & 0 & 0 \\ 
%\twocolor{blue}{purple} 
Others~\tablefootnote{The GitHub ID of nine projects are: vaadin, danfickle, jenkinsci, c2mon, CloudSlang, jitsi, flaxsearch, javadelight, querydsl.} & 9 [9] & 0 & 16 & 0 & 0 & 0 & 0 & 0 & 0 & 0 & 0 & 3 & 2 & 0 & 0 & 0 & 0 & 0 & 0 & 0 \\ 
\hline
\rowcolor{mygray} \textbf{Total} & 43 [64] & 201 & 98 & \textbf{146} & \textbf{26} & 3 & 80 & 0 & 1 & 136 & 121 & 19 & 14 & \textbf{8} & \textbf{9} & 2 & 3 & 0 & 2 & 13 \\

\hline

\end{tabular}

    \label{table:rq1OD}
    % \vspace{-30pt}
\end{table*}

\begin{figure}
    \centering
    \vspace{-20pt}
    \includegraphics[width=0.38\textwidth]{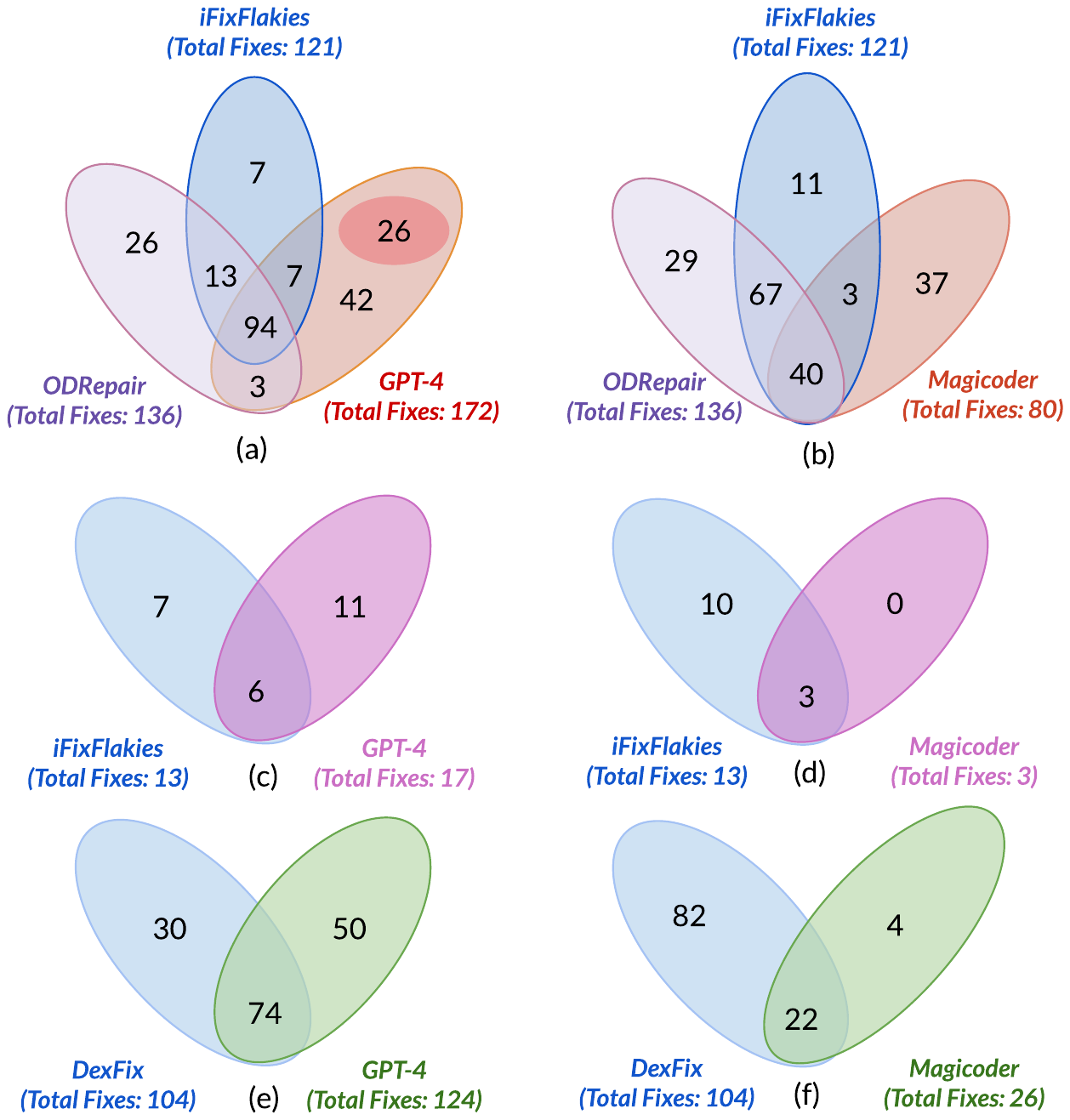}
    \vspace{-20pt}
    \caption{Comparison between the correct patches generated by different approaches. Sub-figures a-b compare OD-Victim, c-d compare OD-Brittle, and e-f compare ID patches}
    \label{fig:venn}
    %\vspace{-10pt}
\end{figure}
% \vspace{-10pt}

\subsection{RQ2: Comparison with Alternative Approaches}
\label{subsec:RQ2}
% \yang{highlight total numbers are higher than others -- added in the figure}
% \yang{PF PU highlights of ifixflakies/odrepair -- this is no need because the "26" is already "previous unfixed" for odrepair}
We further wanted to explore the properties of flaky tests repaired by different approaches. To that end, we illustrate the overlap of successful patches generated by \approach and those from alternative approaches in Figure~\ref{fig:venn}. 
%Figures~\ref{fig:venn}(a) and \ref{fig:venn}(b) show the comparisons of \approach-\gpt and \approach-\magic against \odrepair and \ifixflakies in \odrepair dataset, Figures~\ref{fig:venn}(c) and \ref{fig:venn}(d) present the comparison with \dexfix for ID tests (with a subset of 155 tests from \approach dataset). 
%Given that all patches generated by \magic are subsumed by those from \gpt, we analyzed their effectiveness against other approaches independently. 
For OD-Victim tests (Figures~\ref{fig:venn}a-\ref{fig:venn}b), \approach-\gpt exclusively repairs $68$ OD-Victim tests, including $26$ that were previously unfixed. \approach-\magic can fix 37 tests that neither \odrepair nor \ifixflakies could fix. Alternative OD repair approaches fail at repairing such cases due to the need for existing cleaners (\ifixflakies) or difficulties in identifying complex shared statuses beyond static variables (\odrepair). For OD-Brittle tests (Figures~\ref{fig:venn}c-\ref{fig:venn}d), \approach-\gpt exclusively repairs $11$ of them (nine previously unfixed). \approach-\magic can repair only three OD-Brittle tests, all fixed by \ifixflakies.

Regarding ID tests (Figures~\ref{fig:venn}e-\ref{fig:venn}f), \approach-\gpt successfully repairs $50$ ID flaky tests that \dexfix cannot, and \approach-\magic manages to repair four tests that are beyond the capability of \dexfix, which is limited to specific heuristics and fails to generalize beyond them. Figure~\ref{fig:od-example} and Figure~\ref{fig:id-example} show examples of cases where \approach was able to repair flaky tests, but alternative OD and ID repair approaches could not. 

%In the discussion of unique OD and ID patch examples from \approach in Figure~\ref{fig:od-example} and Figure~\ref{fig:id-example}, we observe the limitations of \ifixflakies which struggle with the absence of existing cleaner tests, and \odrepair has difficulties in identifying complex shared statuses beyond static variables and is hampered by the limitations of test generation tools in generating cleaning statements; while \dexfix faces limitations in repairing ID tests because it is designed on limited heuristics. In this section, we focus on the patches that were successfully fixed by other approaches but not by \approach.

\subsubsection{OD tests that \approach cannot repair} 
%\approach cannot repair $46$ OD-Victim tests and $9$ OD-Brittle tests that alternative approaches do. %There are 46 tests that can not be fixed by \approach, but others can. 
Through manual investigation of cases where \approach could not fix but alternative approaches did, we identified two main recurring patterns: (1) Even though our program analysis provided the polluted variable(s) as context, the patches focused on resetting other variables. In most cases, these variables were directly used in the assertion of the victim method but were not the polluted states. 
(2) Even if LLM identified polluted states correctly, \approach could not generate a correct patch due to hallucination. Examples of such hallucinations include adding variables that do not exist or applying APIs incompatible with the polluted field type. 
%They can be divided into the following reasons: 
%(1) For 26 tests, LLM successfully identified the shared status between polluters and victims.
%, and attempted to reset the states either after the polluter has run or before the victim runs. 
%It generated patches using reset methods with keywords like \texttt{Reset()} or \texttt{ShutDown()}. However, these APIs did not exist for the polluted fields, and LLM did not generate/find the correct reset methods, leading to the failure of these patches. Additionally, these patches always contain compilation errors due to the use of non-existing variables or non-existing types, etc., which remain unresolved in subsequent iterations. 
%(2) For 17 tests, LLMs fail to accurately identify the exact shared status; for example, they focus solely on those methods that explicitly occurred in failure messages and overlook the actual polluted variables.
 %(3) Three of them were generated with false positive patches from \approach; they either surrounded assertions with \texttt{try-catch} or simply removed the assertions in victim to make the test pass. In these cases, the \approach changes the original test logic and fails to resolve the underlying pollution between tests.

\subsubsection{ID tests that \approach cannot repair} There are $30$ tests fixed by \dexfix but not \approach. Breaking down these tests: (1) \approach successfully located the unordered collections but could not generate a correct patch due to overfitting into the provided context. For example, if the assertion failure in the context is related to specific elements in the \texttt{\small{HashMap}}, while LLM creates a \texttt{\small{LinkedHashMap}} (which is an ordered collection), it only populates it with those specific elements and discards others. This may result in resolving the previous assertion failure but failing new ones. (2) \approach successfully located the unordered collections and sorted the elements in a deterministic order. However, it consistently faced compilation errors due to hallucinating unsupported operators or invoking non-existent APIs. 
%Sometimes, LLM also attempts to leverage custom classes from the current project, but it fails to infer the correct uses of them, and finally ends with incorrect API calls.
%(2) For three of them, LLMs were only able to resolve  flakiness partially without identifying the underlying cause. For example, when the flakiness is from complicated third-party libraries or methods created in the main code, LLM struggles to handle such cases due to the lack of context. \reyhan{don't understand this one. will complete after clarification}
%(3) The rest three tests were generated with false positive patches, in which the test method body was fully deleted or rewritten, without addressing the actual source of flakiness,

% \reyhan{add discussion about different approaches complementing each other}
These results confirm that the \approach can complement existing tools for repairing flaky tests.
serve as a complementary technique along with others. For tests where symbolic techniques \dexfix, \ifixflakies, and \odrepair fail to generate a patch based on existing heuristics, developers may use \approach to repair the flakiness.

\subsection{RQ3: Contribution of Different Components}
\label{subsec:RQ3}
In this research question, we evaluate the effectiveness of three notable contributions of \approach: effective bug localization, prompt crafting, and iterative repair. 

%\subsubsection{Effectiveness of Error Message Extraction and Error Code Location}
\subsubsection{Bug Localization}
Without analyzing the test report by the \inspector, \approach should take the entire test execution report of compilation error stack trace. To show the impact of precise bug localization, we sorted all flaky tests based on the length of the original test failure report, selected the top $40\%$ (the budget caps the percentage) ($349$ tests), and replaced the \textit{Failure Location} part of the prompt with the entire test report.
%To show the impact of error extraction in \inspector, we evaluate \approach-\gpt4 and \approach-\magic by not utilizing the extracted error messages and failure test code. Instead, we incorporated the entire test report into the prompt. 
Table~\ref{table:rq3-error-extraction} compares the effectiveness of the \approach with (O-Patches column) or without (Patches column) precise bug localization by the \inspector. These results demonstrate the necessity of minimizing contextual information for LLMs to achieve a higher performance: Without trimming, $435$ prompts (aggregated for both models) exceed the context window of the models. %\reyhan{report accumulated numbers for both models} prompts exceed the context window of the models.
\approach LLMs that originally could repair all selected 
flaky tests only repair $96$ 
of them. 

%171+7+13+189+42+13

%Due to the cost of prompting \gpt, we sorted ID, OD-Victim, and OD-Brittle tests by the length of their test failure reports, selecting the top 40\% with the longest reports for our analysis. Table~\ref{table:rq3-error-extraction} shows successful patch generation in this scenario (S-Patches) alongside the successful patches generated by the original \approach methodology (FD-patches). The results indicate a marked decrease in the success rate of repairs: from $69\%$ with \approach-\gpt to $16\%$, and a reduction from $32\%$ with \approach-\magic to $13\%$, which shows the effectiveness of error extraction in \approach.

\begin{table}[t]
    \setlength{\tabcolsep}{0.5pt}
    \centering
    %\vspace{-10pt}
    % \tiny
    %\scriptsize
    % \footnotesize
    \small
    %\vspace{25pt}
    \caption{Impact of precise bug localization on the effectiveness of \approach. Avg. Lines: Average length of entire test reports; \textbf{Patches} and  \textbf{O-Patches} indicate correct patches with longer prompts and original \approach.
    %\textbf{Avg. Lines}: Average lines of test failure reports; 
    %\textbf{Patches}: Correct patches with longer prompts. 
    %\textbf{O-Patches}: Correct patches from original \approach.
    %\reyhan{please ensure that the text is consistent with the new column names}
    }
    \vspace{-5pt}
    % \footnotesize
\begin{tabular}{|c|c|c|c|c|c|}
\hline
 \textbf{Flakiness} & \textbf{ \#Tests }  
 & \textbf{Avg. Lines}
 & \textbf{Model}
 & \textbf{ \#Patches } & \textbf{ \#O-Patches }  \\
 \hline
 
\multirow{2}{*}{ID} & \multirow{2}{*}{216} &\multirow{2}{*}{606}& 
\gpt & 2 & 111 \\ 
& &  &
\magic & 0 & 28 \\
\hline
\multirow{2}{*}{ OD-Victim} &\multirow{2}{*}{120} &\multirow{2}{*}{164}& 
\gpt  & 52 & 106 \\ 
& &  &
\magic  & 42 & 77 \\
\hline
\multirow{2}{*}{ OD-Brittle} & \multirow{2}{*}{13} &\multirow{2}{*}{489}& 
\gpt  & 0 & 12 \\ 
& & &
\magic  & 0 & 3\\
 \hline
% Total & &  &  &  & \\
\end{tabular}
    \label{table:rq3-error-extraction}
    \vspace{-5pt}
\end{table}

\subsubsection{Prompt Crafting} 
To show the impact of the proposed prompt crafting approach of \approach, we asked \approach-\gpt and \approach-\magic to repair all subject ID and OD flaky tests through \textit{vanilla prompting}: to perform a task without providing additional context. 
Table ~\ref{table:rq3vanilla} shows the result of this experiment. In this experimental setting, \approach-\gpt and \approach-\magic only produced $13$ and two correct patches, compared to $500$ and $170$ original patches. Vanilla prompting results in zero patches for OD-Brittle flaky tests.  
In most of the failed cases, LLM either explicitly mentioned that it does not understand the problem or only explained the test code without producing any patch. \textbf{The huge performance drop ($98\%$) in vanilla prompting indicates the impact of providing the proper context into the prompt.}

\subsubsection{\stitch and Iterative Feedback}
%\yang{showing lifecycle for all patches instead of only ID here}
To investigate the impact of iterative feedback and \stitch, we tracked back the lifetime of patched flaky tests during multiple repair iterations. %\reyhan{we should add some overall numbers and then use the figure to illustrate in more detail how the contribution is happening}
Overall, \textbf{for $500$ flaky tests fixed by \approach-\gpt and $170$ by \approach-\magic, \stitch contributes to $12\%$ and $31\%$ of them, respectively.} 
%To distinguish between different types of flakiness, \hl{14\%} of ID patches and \hl{8\%} of OD patches from \approach-\gpt received contributions from \stitch. Meanwhile, \stitch achieved contribution rates of \hl{57\%} ID and \hl{2\%} OD for \approach-\magic. 
%Figure~\ref{fig:sankey} demonstrates the details of this experiment. %, and additional data are available in our artifact~\cite{website}.  

Figures~\ref{fig:sankey}a-b illustrates the evolution of $170$ and $500$ patches from \approach-\magic and \approach-\gpt. The left grey bar shows the initial state of tests, i.e., being flaky. After applying the patch in each iteration, the status can be \textit{Test Pass (\textbf{P})} (flakiness fixed), \textit{Test Failure (\textbf{F})} (flakiness still exists), or \textit{Compilation Error (\textbf{CE)}} (patching resulted in compilation issue). We are specifically interested in patches that \stitch contributes to changing their status and labeled them with \textbf{[iteration number]:[status1] To [status2]}. For example, ``2:CE To P'' shows \stitch changes the state of patches in iteration 2 from \textit{compilation error} to \textit{test pass}.

Among the $170$ tests successfully repaired by \approach-\magic, \stitch converts compilation errors to \textit{test pass} for $32$ of them (\textbf{CE To P}). For eight tests, while \stitch addressed the compilation errors, the patches resulted in test failures (\textbf{CE To F}). Additionally, for $12$, \stitch resolved partial but not all compilation issues (\textbf{CE To CE}), which also helps to generate improved patches in the subsequent iteration. 
For $500$ correct patches generated by \approach-\gpt, \stitch helped $58$ during the repair process. Among these patches, $37$ were improved by \stitch directly into successful patches. 
%Overall, \stitch contributes to \hl{12\%--31\%} successful patch generation. 
The impact of \stitch is much higher on \magic since it is a smaller and weaker model compared to \gpt. This entails the importance of neuro-symbolic approaches for the ultimate democratization of open-source LLMs.  
%\gpt typically outperforms open-source models across various coding tasks. We believe that the innovative use of the \stitch component can serve as an inspiration for the community, both in the utilization and enhancement of open-source models.

$24\%$ to $60\%$ of patches, including those fixed by \stitch, were generated in the first iteration. The feedback loop contributed to generating the remaining patches in subsequent iterations: \textbf{feedback loop contributes to more than doubling the number of patches generated in the first iteration.}

\begin{table}[t]
    \setlength{\tabcolsep}{0.4pt}
    \centering
    % \tiny
    %\scriptsize
    % \footnotesize
    \small
    %\vspace{25pt}
    \caption{ The results of vanilla prompting compared to \approach.
    %\textbf{\small{Patches}}: Total generated patches; 
    \textbf{Patches}: Total generated patches; 
    \textbf{C-Patches}: Correct patches; 
    \textbf{FP-Patches}: False Positives. 
    %\reyhan{Table is off margin from the top}
    }
    \vspace{-5pt}
    % \footnotesize
\begin{tabular}{|c|c|c|c|c|c|}
\hline
 \textbf{Flakiness} & \textbf{ \#Tests }  & \textbf{Model}
 & \textbf{ \#Patches } & \textbf{ \#C-Patches } & \textbf{ \#FP-Patches }  \\
 \hline
 
\multirow{2}{*}{ ID} &\multirow{2}{*}{541}& 
\gpt & 336 & 11 & 12 \\ 
& & 
\magic & 255 & 1 & 10 \\
\hline
\multirow{2}{*}{ OD-Victim} &\multirow{2}{*}{299}& 
\gpt & 251  & 2 & 2 \\ 
& & 
\magic & 173 & 1 & 2\\
\hline
\multirow{2}{*}{ OD-Brittle} &\multirow{2}{*}{33}& 
\gpt & 13 & 0 & 1 \\ 
& & 
\magic & 11 & 0 & 0\\
 \hline
% Total & &  &  &  & \\
\end{tabular}
    \label{table:rq3vanilla}
    \vspace{-15pt}
\end{table}

%This trend indicates that LLMs are more likely to repair easier patches in the initial rounds, whereas more challenging tests often remain difficult for LLMs to resolve, sometimes resulting in no fix at all

%From left to right, the iteration number increments, indicating the progress of the \approach into iterative repair. \textbf{One consistent observation is that the patches can transform from uncompilable to compilable but with test error to pass as a successful patch.} This is considered a positive evolution of the patch through multiple iterations. We also observe that in some cases, a patch that was previously compilable but with a test failure transformed into a compilation error in the next iteration. While this sounds like a regression, our manual investigation of a subset of such cases revealed that those patches also evolve. That is, \textbf{the changes in the code, although resulting in a compilation error, caused the code to become closer to the patch. In most cases, the subsequent iteration changes the compilation error into pass.} Finally, the large outgoing flow from compilation errors to test failures (or even pass) at each iteration demonstrates the effectiveness of the \textit{Stitching} component. The \textit{Stitching} component can also fix compilation errors, that if not fixed, may require (at least) another round of iteration for the LLM to resolve it. \textit{Stitching} component solves \textit{compilation error} and transforms them into \textit{test failure} or \textit{test pass}, which reduces compilation error by 20\%.

\subsection{RQ4: Performance}
\label{subsec:RQ4}

To address this research question, we evaluated the time and costs involved in using \approach during repairing tests. The iterative workflow of \approach attempts the repair between one to five times.  Tests for which a successful patch is generated may finish earlier, whereas those that cannot be repaired persist longer. On average, \gpt requires 87.2 seconds and costs \$0.12 to repair an ID test or OD-Brittle test, and takes 232.8 seconds at a cost of \$0.27 to complete a repair attempt for unsuccessful tests. For OD tests, \gpt needs 107.5 seconds at a cost of \$0.18 to repair a test successfully, and 214.2 seconds costing \$0.35 for unsuccessful repair attempts. \magic, on the other hand, takes 109.2 seconds to successfully repair an ID test and 355.9 seconds for an unsuccessful attempt; for OD tests, it requires 110.6 seconds for a successful repair and 247.7 seconds for an unsuccessful attempt.

\begin{figure}[t]
    \centering
    %\vspace{-5pt}
    \includegraphics[width=0.48\textwidth]{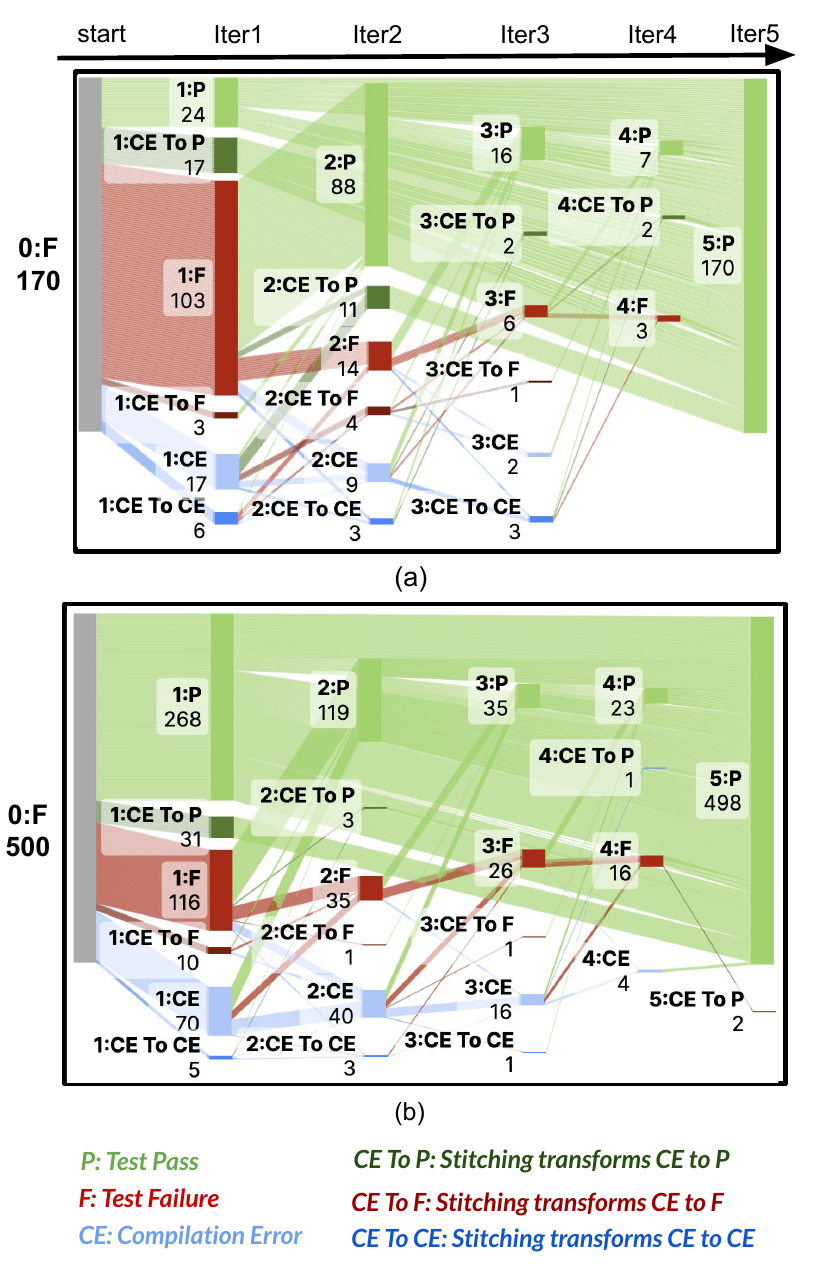}
    \vspace{-10pt}
    \caption{ The evolution of patches through different repair iterations of (a) \approach-\magic and (b) \approach-\gpt. 
    %(\textbf{{\small{P}}}: \textit{Test Pass};  \textbf{\small{F}}: \textit{Test Failure}; \textbf{\small{CE}}: \textit{Compilation Error}; 
    The notation \textit{\textbf{To}} indicates the applicability and impact of the \stitch to the current patch
    }
    \label{fig:sankey}
    %\vspace{-10pt}
\end{figure}
\section{Related Work}
\label{sec:relatedwork}

%\cite{zhang2021domain}
%Test flakiness has become a hot research topic in the software engineering community over the past few years. Several empirical studies confirm the importance of this problem from the developer's perspective~\cite{luo2014empirical,eck2019understanding}. 
Many techniques have been proposed for characterising~\cite{lam2019root,lam2020study,lam2020large,lam2020understanding,dutta2020detecting}, detecting~\cite{ziftci2020flake,wang2022ipflakies,person2015test,mascheroni2018identifying,bell2018deflaker,king2018towards,lam2019idflakies,pinto2020vocabulary,verdecchia2021know,yi2021finding,wei2021probabilistic}, or repairing~\cite{dutta2021flex,li2022repairing,pei2023traf,shi2019ifixflakies,wang2022ipflakies,wei2022preempting} test flakiness. Recently, Chen et al. proposed Croissant~\cite{chen2023transforming}, a tool for modifying tests such that a non-flaky test suite shows flaky behavior. \ifixflakies~\cite{shi2019ifixflakies} and \ipflakies~\cite{wang2022ipflakies} are two related research on repairing test flakiness exist in Java and Python tests suites. %, respectively. 
%To repair OD flaky tests in the test suite of Java projects, 
\ifixflakies takes the order-dependent test, the failing test order, and the passing test order. The current implementation of \ifixflakies leverages \idflakies to get the required inputs. 
%relies on \idflakies to identify victim OD tests. 
%\yang{maybe we don't say ifixflakies relies on idflakies here, which seems like ifixflakies integrated idflakies, we can say:iFixFlakies takes as input an order-dependent test, a passing test order, and a failing test order, the test orders can be generated from idflakies.} 
It then modifies the execution order of different sub-sequences of tests to find tests that modify the shared state---by setting or unsetting the shared states---with the identified victim or brittle, and uses them to generate the patch. \ipflakies follows similar steps but can only repair victim OD tests (not brittles) in Python test suites. Compared to these approaches, \approach is more versatile in repairing both victim and brittle OD tests as well as ID flaky tests. 

\odrepair~\cite{li2022repairing} is proposed to overcome the limitation of \ifixflakies, which rely on the existence of \textit{cleaner} tests to repair victim OD tests. To that end, it analyzes the static fields and serialized heap state to identify the polluted shared states between victim and polluter tests and relies on automated testing techniques to generate cleaners tests. By enforcing the execution of cleaner tests before the victim, \odrepair resolves the test flakiness. Compared to this technique, which only targets repairing victim OD tests, \approach can repair more categories of test flakiness. Also, our proposed technique completely resolves the dependency, making the patch more realistic to resolve test flakiness. 

\dexfix~\cite{zhang2021domain} repairs ID flakiness by implementing domain-specific repair strategies that resolve implementation dependencies in both the test and the main codes. 
%Despite the effectiveness of \dexfix on repairing ID flaky tests, 
%repairing $119$ out of $275$ ID flaky tests, 
Consequently, it is limited to strategies tailored to repair studied flaky tests and may not generalize to other patterns. \approach is not limited in that way due to relying on LLMs to perceive the nature of test flakiness and repairing based on the relevant contexts provided in the prompt. 

TRaF~\cite{pei2023traf} aims to address test flakiness in the JavaScript test suite of web-based applications by updating the waiting time of asynchronous calls to a value that breaks the time dependency between tests. To that end, they use code similarity and look at the relevant change history of the code, hoping to find useful hints for the efficient wait time in the existing or past code versions. Asynchronous waits are a subcategory of NOD test flakiness~\cite{chen2023transforming}, which the current implementation of \approach does not support. 
%While we have not evaluated \approach on JavaScript tests, we anticipate it can perform relatively well in repairing OD and ID JavaScript flaky tests due to generalizing GPT-4 to different programming languages~\cite{gpt4}. In such a case, \approach will be orthogonal to TRaF to repair OD and ID flaky tests. 

In addition to the mentioned differences with each related work, \approach is the first test flakiness repair technique that leverages the power of LLMs. The empirical evaluation clearly shows the benefit of using LLMs, i.e., repairing different categories of test flakiness and generating successful patches for flaky tests that were not previously fixed by humans or existing automated techniques.  

\vspace{-5pt}
\section{Concluding Remarks}
\label{sec:conclusion}

%Test Flakiness is an important topic in software testing and analysis, not only in academia but also in the industry. 
%Despite the importance of test flakiness problem, flakiness detection and repair techniques are still in their infancy. Specifically, t
%There is a dearth of research on automated repair of different types of test flakiness. 
In this paper, we proposed \approach, the first technique that combines the generalizability power of LLMs with the soundness of the program analysis, to repair different types of flakiness. Our evaluation results show that \approach is able to generate patches for flaky tests of real-world projects that were previously unfixed. In many cases, neither prior automated techniques nor human developers were able to repair such flakiness. 

We are considering several research directions on top of this work. The first obvious plan is supporting the repair of NOD flaky tests. This requires devising more complex analysis techniques to localize such flakiness issues and revising the prompt template to incorporate relevant context. Next, we plan to perform a large-scale empirical study to further pinpoint when \approach can repair test flakiness, and when it cannot. This would provide insight into the research gap, very likely to require more advanced offline processing techniques to further help LLMs repair flaky tests. 

% \section{Data Availability}

% Our artifact is publicly available at ~\cite{website}.
\vspace{5pt}

\bibliographystyle{ACM-Reference-Format}
\bibliography{refs.bib}
% \addbibresource{refs.bib}

\end{document}